\documentclass{article}
\usepackage{placeins}
\usepackage{amsmath}
\usepackage{microtype}
\usepackage{graphicx}
\usepackage{subcaption}
\usepackage{booktabs} 
\usepackage{makecell}

\usepackage{hyperref}
\usepackage{multirow}
\usepackage[table,xcdraw]{xcolor}
\usepackage{colortbl}

\usepackage{amsmath}
\usepackage{amssymb}
\usepackage{mathtools}
\usepackage{amsthm}

\usepackage{algorithm}
\usepackage{algorithmic}

\usepackage[capitalize,noabbrev]{cleveref}
\usepackage{arxiv}

\usepackage[utf8]{inputenc} 
\usepackage[T1]{fontenc}    
\usepackage{hyperref}       
\usepackage{url}            
\usepackage{booktabs}       
\usepackage{amsfonts}       
\usepackage{nicefrac}       
\usepackage{microtype}      
\usepackage{lipsum}
\usepackage{graphicx}
\graphicspath{ {./images/} }

\title{Beyond Conditional Computation: Retrieval-Augmented Genomic Foundation Models with Gengram}

\author{
 Huinan Xu \\
  Genos Team\\
 \\
   \And
 Xuyang Feng \\
  Genos Team\\
 \\
   \And
 Junhong Chen \\
  Genos Team\\
 \\
   \And
 Junchen Liu \\
  Genos Team\\
 \\
   \And
 Kaiwen Deng \\
  Genos Team\\
 \\
   \And
 Kai Ding \\
  Genos Team\\
 \\
   \And
 Shengning Long \\
  Genos Team\\
 \\
    \And
 Jiaxue Shuai \\
  Genos Team\\
 \\
   \And
 Zhaorong Li \\
  Genos Team\\
 \\
   \And
 Shiping Liu \\
  Genos Team\\
 \\
   \And
 Guirong Xue \\
  Genos Team\\
 \\
   \And
 Zhan Xiao \\
  Genos Team\\
  \texttt{xz@zhejianglab.org}\\
}

\begin{document}
\maketitle
\begin{abstract}
    Current genomic foundation models (GFMs) rely on extensive neural computation to implicitly approximate conserved biological motifs from single-nucleotide inputs. We propose Gengram, a conditional memory module that introduces an explicit and highly efficient lookup primitive for multi-base motifs via a genomic-specific hashing scheme, establishing genomic "syntax". Integrated into the backbone of state-of-the-art GFMs, Gengram achieves substantial gains (up to 14\%) across several functional genomics tasks. The module demonstrates robust architectural generalization, while further inspection of Gengram’s latent space reveals the emergence of meaningful representations that align closely with fundamental biological knowledge. By establishing structured motif memory as a modeling primitive, Gengram simultaneously boosts empirical performance and mechanistic interpretability, providing a scalable and biology-aligned pathway for the next generation of GFMs. The code is available at \url{https://github.com/zhejianglab/Genos}, and the model checkpoint is available at \url{https://huggingface.co/ZhejiangLab/Gengram}. 
\end{abstract}


\section{Introduction}
DNA sequences constitute the fundamental biological language of life, encoding the genetic information underlying cellular function and organismal development. Understanding the “syntax” of this language is central to deciphering biological processes and their dysfunctions \cite{kernohan2024expanding,yang2025regulatory}. Recent advances in GFMs on architectures and tokenizers (e.g., k-mer, Byte Pair Encoding(BPE), single-base) have substantially accelerated progress in this direction.

Although single-nucleotide tokenization has demonstrated strong performance in downstream tasks, particularly those that evaluate fine-grained single-base effects \cite{lin2025genos}, the majority of biologically functional genomic elements are defined by multi-nucleotide motifs with specific sequence patterns. Consequently, constrained by the Transformer architecture, existing Transformer-based GFMs lack an explicit mechanism to store and retrieve such functional motifs, instead relying on large-scale pretraining and dense computation to implicitly infer them from combinations of single-base representations \cite{tzanakakis2026fundamental}. This indirect inference process is inefficient and often limits performance on motif-dominated functional element detection tasks.

Inspired by DeepSeek's Engram—a conditional memory module that stores n-gram knowledge in a hash-based memory \cite{cheng2026conditional} — we propose Gengram, a conditional memory module specifically designed for genomic motif modeling. In Gengram, k-mers (with k = 1,2,3,4,5,6) and their corresponding embeddings are explicitly stored and leveraged through a gate-controlled conditional memory mechanism. Unlike Engram, which retrieves single preceding n-grams per position, Gengram introduces a \textbf{local window aggregation mechanism}, which aggregates multiple k-mer embeddings within a fixed genomic window. This design allows the model to capture local contextual dependencies relevant to functional motifs while avoiding reliance solely on single-nucleotide composition. 

Owing to the small vocabulary size of genomic sequences (ATCGN), Gengram introduces negligible additional computational overhead while enabling larger and more expressive motif memory tables. More importantly, Gengram can be seamlessly integrated into mainstream Transformer architectures. Our contributions can be summarized as follows:

\textbf{We propose Gengram}, a lightweight conditional motif memory module for Transformer-based GFMs that encodes multi-nucleotide motifs and integrates them via a structure-aware windowed aggregation mechanism, improving efficiency and interpretability without hard-coded biological rules.

\textbf{Gengram demonstrates consistent and substantial performance improvements}  across downstream genomic tasks—especially motif-dominated functional element detection—achieving up to 14\% improvement over state-of-the-art GFMs architectures under identical training protocols. 

\textbf{Gengram is architecture-agnostic,}  introducing negligible additional parameters (\textasciitilde 60M) and computation, while consistently reducing training loss across diverse Transformer architectures, suggesting its potential as a reusable, standardized component for future GFMs.

\textbf{Gengram is computationally efficient,} achieving benchmark performance comparable to state-of-the-art models while utilizing fewer activated parameters and less training data. The Gengram module exhibits robust capability in modeling long sequences and enables stable load balancing in sparse Mixture of Experts (MoE) models.

\textbf{Gengram spontaneously captures biologically meaningful structure,} exhibiting reverse-complement symmetry in memory embeddings and context-dependent gating aligned with functional genomic regions (e.g., promoters and 5$^{\prime}$ UTRs), indicating structural reasoning rather than static pattern memorization.

\section{Related Works}
\subsection{Current Architectures and Representation Paradigms in GFMs }
\subsubsection{Genomic Foundation Models}
Current genomic modeling is dominated by Transformer-derived architectures \cite{vaswani2017attention} and non-Transformer alternatives, primarily represented by State Space Models (SSMs) \cite{gu2024mamba}. Based on their underlying mechanisms, these models can be categorized into three paradigms: Bidirectional Encoders (e.g., DNABERT) \cite{ji2021dnabert}: Built upon the Transformer Encoder, these models utilize omnidirectional attention for local motif extraction. Autoregressive Decoders (e.g., Nucleotide Transformer) \cite{dalla2025nucleotide,boshar2025foundational}: Leveraging the Transformer Decoder, these models offer robust long-range modeling capabilities. State Space Models (e.g., Caduceus) \cite{schiff2403caduceus}: Based on the Mamba architecture, SSMs achieve linear scaling via selective scanning.

\section{Method}
\begin{figure*}[ht]
  \vskip 0.2in
  \begin{center}
    \centerline{\includegraphics[width=\textwidth]{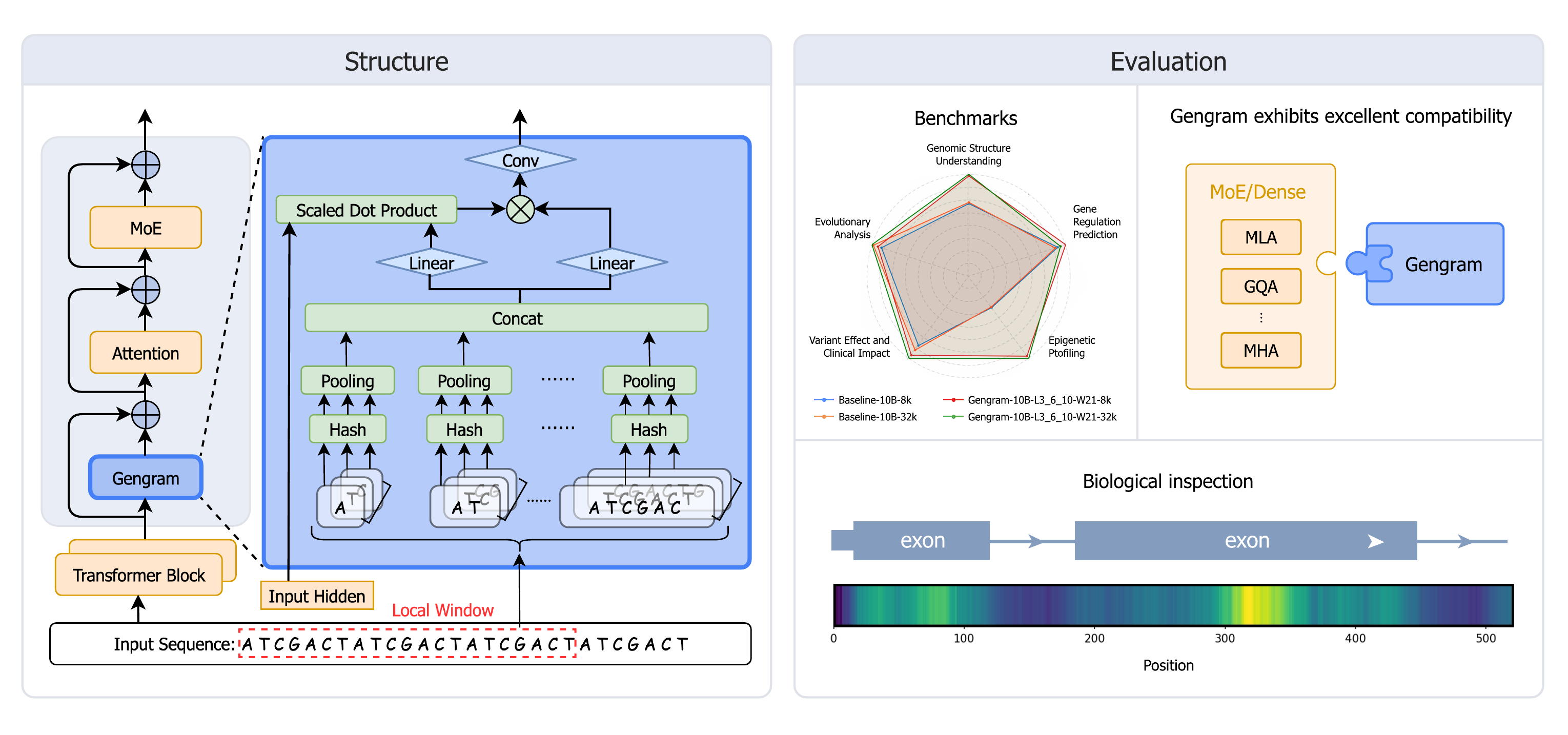}}
    \caption{
      \textbf{Overview of the Gengram architecture and evaluation.} (Left) Gengram is integrated into a Transformer block via a residual connection and applied before the attention module. It maintains a motif memory implemented as a hash table with static keys derived from all possible k-mers ( k=1,2,3,4,5,6) and learnable embedding values. For a given local window, k-mers are mapped to memory entries, aggregated within each k-mer level, and concatenated across levels. The resulting representation is then passed through a gate-controlled module to modulate the hidden states. (Right) Evaluation and analysis of Gengram. We conduct extensive benchmarks across diverse genomic tasks and Transformer architectures, demonstrating strong compatibility with different attention mechanisms (e.g., MHA, GQA, MLA). Biological inspection further shows that Gengram’s activations align with functional genomic regions.
    }
    \label{fig:fig1}
  \end{center}
\end{figure*}
\subsection{Architecture Overview}
As shown in Figure \ref{fig:fig1}, the Gengram module integrates efficient lookup with a motif prior: when predicting the current token, it uses a sliding-window mechanism within the Local Window immediately preceding it to retrieve all occurring contiguous substrings, i.e., motif priors. After deduplication, it looks up the corresponding values in a hash table; vectors with the same $N$ are summed and averaged, vectors across different $N$ are concatenated, and then, after a linear transformation, dot-product computation, and gating, the resulting signal is added to the backbone network via a residual connection.
\subsection{Notation}
Let the backbone produce an input representation and the previous-layer hidden states are
\begin{equation}
    X \in \mathbb{R}^{n \times d_{\text {model }}},X^{(\ell-1)} \in \mathbb{R}^{n \times d_{\text {model}}} 
    \label{equ1}
\end{equation}
where $n$ is the sequence length, $d_{\text {model}}$ the hidden dimension, and $X_{\text {t}}$. 

The $N$-gram lengths follow chapter \ref{c4.2}
\begin{equation}
    N \in \mathcal{N}=\{1,2,3,4,5,6\}
    \label{equ3}
\end{equation}

To capture multi-scale motifs. For each $N$, we maintain a learnable lookup hash table
\begin{equation}
   \left\{\mathbf{M}^{(N)}\right\}_{N \in \mathcal{N}}, \quad \mathbf{M}^{(N)} \in \mathbb{R}^{|\Sigma|^{N} \times d_{m}}
   \label{equ4}
\end{equation}

where $\Sigma$ is the token alphabet (A,T,C,G,N) and $d_{m}$ is the memory vector dimension. Intuitively, $\mathbf{M}^{(N)}$ stores an embedding for each possible $N$-gram key; the key-space size is $|\Sigma|^{N}$. The mapping from an $N$-gram $g \in \Sigma^N$ to a row index is defined by a hash function
\begin{equation}
    h_{N}(\cdot) : \Sigma^N\rightarrow \{0,1,...,|\Sigma|^N-1\}
    \label{equ5}
\end{equation}

Since $|\Sigma|$ is small, $h_{N}(\cdot)$ can use a deterministic base-$|\Sigma|$ encoding, giving efficient collision-free indexing.
\subsection{Local Window and N-gram Enumeration}
For each position t, we define a causal window index set
\begin{equation}
    \mathcal{W}_{t}=\{t-W, t-W+1, \ldots, t-1\}
    \label{equ6}
\end{equation}

We scan $N$-grams over the local window $\mathcal{W}_{t}$ to obtain a complete set of substrings, enabling the model to capture multi-scale motif information. The local window also reduces noise and supports efficient retrieval.

Within this window, we enumerate all contiguous $N$-grams using a sliding window scan, as shown in Algorithm \ref{alg:retrieve_at_position}: 

\begin{algorithm}[tb]
  \caption{Sliding window search}
  \label{alg:retrieve_at_position}
  \begin{algorithmic}
    \STATE Retrieve at Position $t$
    \STATE {\bfseries Input:} position $t$
    \STATE {\bfseries Output:} $\{ m_t^{(N)} \}_{N \in \mathcal{N}}$
    \FOR{$N \in \mathcal{N}$}
      \STATE $j_{start} \gets \max(1,\, t - W)$
      \STATE $j_{end} \gets t - N$
      \STATE $sum \gets \mathbf{0}_{d_m}$ \hfill {\itshape initialize accumulator}
      \STATE $cnt \gets 0$ \hfill {\itshape count valid $N$-grams}
      \FOR{$j = j_{start}$ {\bfseries to} $j_{end}$}
        \STATE $key \gets h_N(s[j \ldots j+N-1])$ \hfill {\itshape build $N$-gram key}
        \STATE $sum \gets sum + M^{(N)}[key]$ \hfill {\itshape $O(1)$ lookup}
        \STATE $cnt \gets cnt + 1$
      \ENDFOR
      \STATE $m_t^{(N)} \gets sum / \max(cnt, 1)$ \hfill {\itshape mean pooling}
    \ENDFOR
    \STATE {\bfseries return} $\{ m_t^{(N)} \}_{N \in \mathcal{N}}$
  \end{algorithmic}
\end{algorithm}

\begin{equation}
    \mathcal{G}_{t}^{(N)}=\left\{\left(s_{j},  \ldots, s_{j+N-1}\right) \mid j \in[t-W, t-N]\right\}
    \label{equ7}
\end{equation}

where $s_j$ is the discrete token at position $j$. To make the aggregation focus on distinct motif types rather than their raw frequency, we first deduplicate the retrieved substrings:
\begin{equation}
    \tilde{\mathcal{G}}_{t}^{(N)}=\operatorname{Unique}\left(\mathcal{G}_{t}^{(N)}\right)
    \label{equ8}
\end{equation}

\subsection{Hash Retrieval and Within-$N$ Aggregation}
Each $N$-gram $g \in \Sigma^N$ is retrieved from the correspoding table by
\begin{equation}
    v(g)=\mathbf{M}^{(N)}\left[h_{N}(g)\right], \quad g \in \tilde{\mathcal{G}}_{t}^{(N)}
    \label{equ9}
\end{equation}

We then average the retrieved vectors to obtain a fixed-size summary for length $N$: 
\begin{equation}
    m _ { t } ^ { ( N ) } = \frac { 1 } { | \tilde { \mathcal { G } } _ { t } ^ { ( N ) } | } \sum _ { g \in \tilde { \mathcal { G } } _ { t } ^ { ( N ) } } \mathbf { M } ^ { ( N ) } [ h _ { N } ( g ) ]
    \label{equ10}
\end{equation}

We apply mean pooling to compress a variable number of retrieved $N$-gram values into a fixed-dimensional summary $m _ { t } ^ { ( N ) }$. This choice stabilizes the magnitude of $m _ { t } ^ { ( N ) }$across positions and sequences. Such scale stability makes the contribution of the memory branch easier to control and more robust during training.

\subsection{Cross-$N$ Fusion }
We fuse multi-scale summaries by concatenation: 
\begin{equation}
    m_{t} = \mathrm{Concat}(m_t^{(1)},m_t^{(2)},m_t^{(3)},m_t^{(4)},m_t^{(5)},m_t^{(6)})
    \label{equ11}
\end{equation}

We then apply two linear projections to the same $m_{t}$:
\begin{equation}
    z_t = m_t W_z +b_z,\ u_t = m_tW_u+b_u
    \label{equ12}
\end{equation}

Here $z_t$ is used to compute a gate conditioned on the token representation, while $u_t$ carries the memory content to be injected. 

\subsection{Scaled Dot-product Gating}
We compute a gating scalar via a scaled dot product between normalized vectors: 
\begin{equation}
   \tilde{u}_{t} = \sigma\!\left(\frac{\mathrm{RMSNorm}(X_t)^{\mathrm {T}}\mathrm{RMSNorm}(z_t)}{\sqrt{d_{\text{model}}}}\right)u_t
   \label{equ13}
\end{equation}

We compute a gating scalar from the scaled dot product of the normalized input $X_t$ and normalized projection $z_t$, where $\sigma(\cdot)$ is the sigmoid function and RMSNorm is root mean square normalization, to control how strongly the retrieved motif memory affects the backbone at each position; it is then dotted with $u_t$.

Then we align and mix the gated memory signal through a lightweight transformation, and add it residually to the previous-layer hidden state:
\begin{equation}
    Y_{t} = \mathrm {SiLU}( (\mathrm {RMSNorm}(\tilde{u}_{t} ))) + X_{t}^{(\ell -1)}
    \label{equ15}
\end{equation}

The resulting $ Y \in \mathbb{R}^{n\times d_{\text{model}}}$is fed to subsequent network components. This residual formulation stabilizes training while allowing the model to selectively incorporate explicit motif-level information from the windowed context.


\subsection{Time Comlexity (shows as $O(n)$)}
For each position $t$ and each $N \in \mathcal{N}$, we enumerate approximately max$(0,W-N+1)$ contiguous $N$-grams within the lock-back window and perform one table lookup ($O(1)$) per $N$-gram. Therefore, the total retrieval cost over a length-n sequence is:
\begin{equation}
    O\left(n \sum_{N \in \mathcal{N}} \max (0, W-N+1)\right)=O(n|\mathcal{N}| W)
    \label{equ16}
\end{equation}

When $W$ and $\mathcal{N}$ are treated as fixed hyperparameters (constants), the retrieval time is linear in the sequence length: $O(n)$.

\section{Experimental Design}
\subsection{Datasets}
\subsubsection{Pre-training Dataset} 
The pre-training dataset is curated from the Human Pangenome Reference Consortium (HPRC, release 2) \cite{liao2023draft} and NCBI RefSeq databases \cite{goldfarb2025ncbi}, encompassing DNA sequences from humans and various primate species. To support different stages of model development, we systematically constructed three datasets: (1) a 50B-token dataset with a sequence length of 8,192, designed for the ablation studies; (2) a 200B-token dataset with an 8k context length; and (3) a 100B-token dataset with a 32k context length, both designated for the formal pre-training of the 10B model. Across all datasets, a balanced 1:1 ratio between human and non-human sequences was strictly maintained to ensure taxonomic parity. Detailed list of the datasets are provided in the Appendix Table \ref{atable:train_dataset_1.2B8k},\ref{atable:train_dataset_10B8k} and \ref{atable:train_dataset_10B32k}.

\subsubsection{Benchmark Datasets} 
We utilized several standard benchmark datasets to evaluate the model’s performance, including the Genomic Benchmarks (GB), the Nucleotide Transformer Benchmarks (NTB), the Long-Range Benchmarks (LRB), and the Genos Benchmarks (GeB).\cite{grevsova2023genomic,dalla2025nucleotide,trop2024genomics,lin2025genos} From these sources, we selected 18 datasets covering five major categories: Genomic Structure Understanding, Gene Regulation Prediction, Epigenetic Profiling, Variant Effect and Clinical Impact, and Evolutionary Analysis. Detailed descriptions of each dataset are provided in the Appendix Table \ref{atable:Benchmarks Datases}. All tasks follow a zero-shot evaluation protocol, where the pre-trained model serves as a frozen backbone for feature extraction. Downstream predictions are evaluated via Multi-Layer Perceptron (MLP) and XGBoost classifiers to quantify the model's zero-shot performance.

\subsection{Multi-scale K-mers} \label{c4.2}
The selection of multi-scale k-mers is motivated by both biological and computational considerations. Biologically, different functional genomic elements are characterized by sequence patterns of varying lengths: splice sites are primarily determined by short dinucleotide signals \cite{mount1982catalogue}, codons encoding amino acids consist of triplets \cite{crick1968origin}, while other regulatory elements—such as promoter motifs (e.g., the TATA box)—span longer sequence segments \cite{smale2003rna}. Computationally, we restrict k-mers to lengths up to 6 based on the efficiency–precision trade-off validated in GENERator \cite{wu2025generator}. 

\subsection{Gengram Layer Selection Experiments}
Gengram, as a conditional memory module, only needs to be inserted into a small subset of Transformer layers \cite{cheng2026conditional}. In our 1.2B-parameter model with 0.3B activated parameters (configuration details are provided in Appendix Table \ref{atable:1.2b model structure}), we vary only the insertion layer of Gengram and compare the validation loss on a unified validation set, thereby selecting the optimal single-layer placement that achieves the lowest validation loss. 

\begin{figure}[ht]
  \vskip 0.2in
  \begin{center}
    \centerline{\includegraphics[width=0.5\columnwidth]{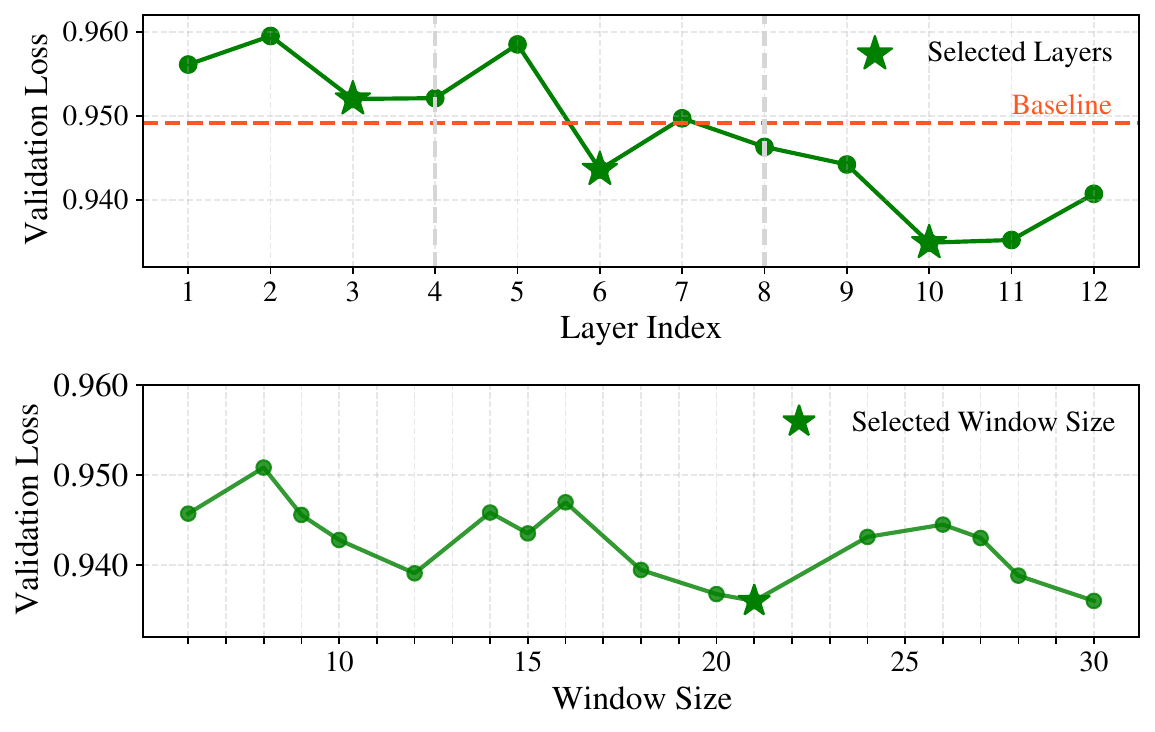}}
    \caption{ \textbf{Parameter Selection Experiments.} 
      \textbf{Top:} Validation loss when inserting Gengram into individual Transformer layers in a 1.2B-parameter (0.3B activated) model trained on 50B tokens. The red curve denotes the baseline without Gengram. \textbf{Bottom:} Validation loss under different Gengram window sizes.
    }
    \label{fig:selected_layer_window}
  \end{center}
\end{figure}

As shown in Figure \ref{fig:selected_layer_window} top, the layer-sweep results indicate that, as Gengram is inserted into progressively deeper layers, the validation loss decreases significantly. The reason is that, in shallow layers, the backbone hidden states are more biased toward local morphology and short-range statistical representations and are sensitive to base-/short-fragment patterns \cite{jawahar2019does}, which can easily lead to over-retrieval; however, shallow injection makes the model less likely to miss truly useful motif evidence. In deep layers, after multiple rounds of attention aggregation, the deep hidden states have already integrated longer-range dependencies, positional relationships, and information beyond the upstream window, and the representations on which the gating decision relies are more informative and more aligned with the task objective, enabling better discrimination between functional motifs and non-functional repeats, thereby yielding lower validation loss \cite{vig2019analyzing,avsec2021effective}.

Validation loss from single-layer insertion alone is insufficient to conclude. Because shallow and deep layers are complementary, we evaluate multi-layer combinations. Prior studies commonly partition Transformer depth into shallow, middle, and deep stages: shallow layers mainly form the basic representational space, middle layers emphasize abstraction and compression, and deep layers exhibit stronger reasoning capability and task specialization \cite{skean2025layer}. Following this view and inspired by the hierarchical processing of the cerebral cortex—transitioning from basic feature extraction in posterior sensory regions (corresponding to shallow layers) and information integration in medial regions (middle layers) to high-level cognitive decision-making in anterior regions (deep layers)—we select the best-performing layer from each stage (namely layers \{3, 6, 10\}) as a representative shallow/middle/deep trio \cite{margulies2016situating}. This combination effectively captures the model's full trajectory from base-level detail scanning to high-level functional abstraction.  In addition, guided by the single-layer rankings, we test several competitive combinations, including \{10, 11, 12\}, \{9, 10, 11, 12\}, \{3, 6\}, and \{3, 10\}. We assess multi-layer benefits with a suite of downstream evaluations. As reported in Table \ref{table:Select_layers}, we summarize the average performance across five downstream evaluation categories for single-layer versus multi-layer insertions. The \{3, 6, 10\ configuration performs best overall, ranking first in four out of five categories.
\begin{table*}[t]
  \caption{\textbf{Downstream Evaluation of Multi-Layer:} Average performance across five downstream categories for single-layer vs. multi-layer insertions. {3, 6, 10} performs best overall, ranking first in four of five categories.}
  \label{table:Select_layers}
  \begin{center}
    \begin{small}
      \begin{sc}
        \begin{tabular}{lcccccc} 
          \toprule
          Selected Layers                   & 10                    & 10,11,12 & 9,10,11,12 & 3,6,10                & 3,6    & 3,10   \\
          \midrule
        Genomic Structure Understanding   & \textbf{0.8944} & 0.8866   & 0.8900     & 0.8937          & 0.8838 & 0.8883 \\
        Gene Regulation Prediction        & 0.8206          & 0.8154   & 0.8188     & \textbf{0.8222} & 0.8181 & 0.8196 \\
        Epigenetic Profiling              & \textbf{0.8334} & 0.8329   & 0.8333     & \textbf{0.8334} & 0.8276 & 0.8306 \\
        Variant Effect \& Clinical Impact & 0.7577          & 0.7553   & 0.7489     & \textbf{0.7678} & 0.7602 & 0.7649 \\
        Evolutionary Analysis             & 0.9101          & 0.9049   & 0.9059     & \textbf{0.9117} & 0.9093 & 0.9064 \\
          \bottomrule
        \end{tabular}
      \end{sc}
    \end{small}
  \end{center}
  \vskip -0.1in
\end{table*}

\subsection{Window Size Selection Experiments}
Using the 1.2B parameter model’s training setup (see  Appendix Table \ref{atable:1.2b model structure}), we determined the optimal window length for Gengram with multiples of 2 and 3. Keeping the backbone, data, training budget, and all training hyperparameters—such as sequence length, batch size, optimizer, and learning rate—exactly the same, and inserting the Gengram module at the tenth layer based on preliminary single-layer experiments, we trained models with different Gengram window lengths under the same token/step budget (i.e., 50B tokens). We then compared the validation loss on a fixed validation set to select the window length that achieved the lowest validation loss. This optimal window length was subsequently used as the baseline for follow-up experiments.

By comparing the validation loss under different window sizes, as shown in Figure \ref{fig:selected_layer_window} bottom, we found that the loss reached its minimum when the window size was set to approximately 21. This optimal length coincides with the fundamental structural property of B-form DNA, the predominant natural DNA: it completes one helical turn approximately every 10.4~10.5 base pairs (bp) in cellular environments\cite{wang1979helical}. Consequently, base pairs separated by every two turns (i.e., approximately 21 bp) are positioned in the same face of the helix \cite{chargaff1950chemical}. This spatial arrangement has a significant impact on functions related to transcriptional regulation, particularly those involving protein binding. This structural consistency provides a universal constraint across the entire genome, including non-coding regions \cite{luger1997crystal}. By explicitly modeling this 21-bp periodicity, our model effectively captures these phasing correlations, allowing it to leverage the stereochemical patterns that recur with every two helical turns—such as mutation susceptibility or protein-binding preferences. Therefore, the Gengram module moves beyond local linear context to incorporate global structural priors, significantly improving the performance of whole-genome modeling \cite{wang1996dna}.

\subsection{Pretraining Setup}
To validate the effectiveness of the proposed Gengram module, we conducted experiments by training a 10B-parameter model named Gengram-10B-L3\_6\_10-W21-8k using the Megatron-LM framework on a distributed system with 8 GPUs per node and pipeline parallelism of 2. The model architecture consisted of 12 transformer layers with a hidden size of 4096 and 16 attention heads, incorporating RMSNorm, Rotary Position Embedding (RoPE), Switched Gated Linear Unit (SwiGLU) activation, and group query attention with 8 query groups (see Appendix Table \ref{atable:w-wo Gengram}). The Gengram module was enabled at layers 3, 6, and 10, configured with n-gram sizes from 1 to 6, an embedding dimension of 1024 per n-gram, token IDs (8, 5, 6, 7, 9) representing nucleotides A, C, G, T, and N respectively, a short convolution of kernel size 4, and a local window size of 21. The design also employed MoE architecture with 8 experts, top‑$k$ routing ($k$=2), auxiliary loss-based load balancing, and Grouped GEMM (General Matrix Multiplication) operations. The model was trained on a mixed human and non‑human genomic dataset comprising approximately 25 million samples (200B tokens), with a sequence length of 8192 and maximum position embeddings of 8192. Key hyperparameters included a global batch size of 1024, an initial learning rate of 1e‑4 decaying cosine to 1e‑5, weight decay of 0.1, gradient clipping at 1.0, and BF16 precision with flash attention. To ablate the contribution of the local window mechanism, a variant without the window, Gengram‑10B‑L3\_6\_10‑8k, was also trained. Additionally, a baseline model without the Gengram module, Baseline‑10B‑8k, was trained for comparison.
\section{Results \& Analysis}
\subsection{Enhanced Performance of the Gengram Module and the Role of the Window Mechanism}
\begin{figure*}[ht]
  \vskip 0.2in
  \begin{center}
    \centerline{\includegraphics[width=\textwidth]{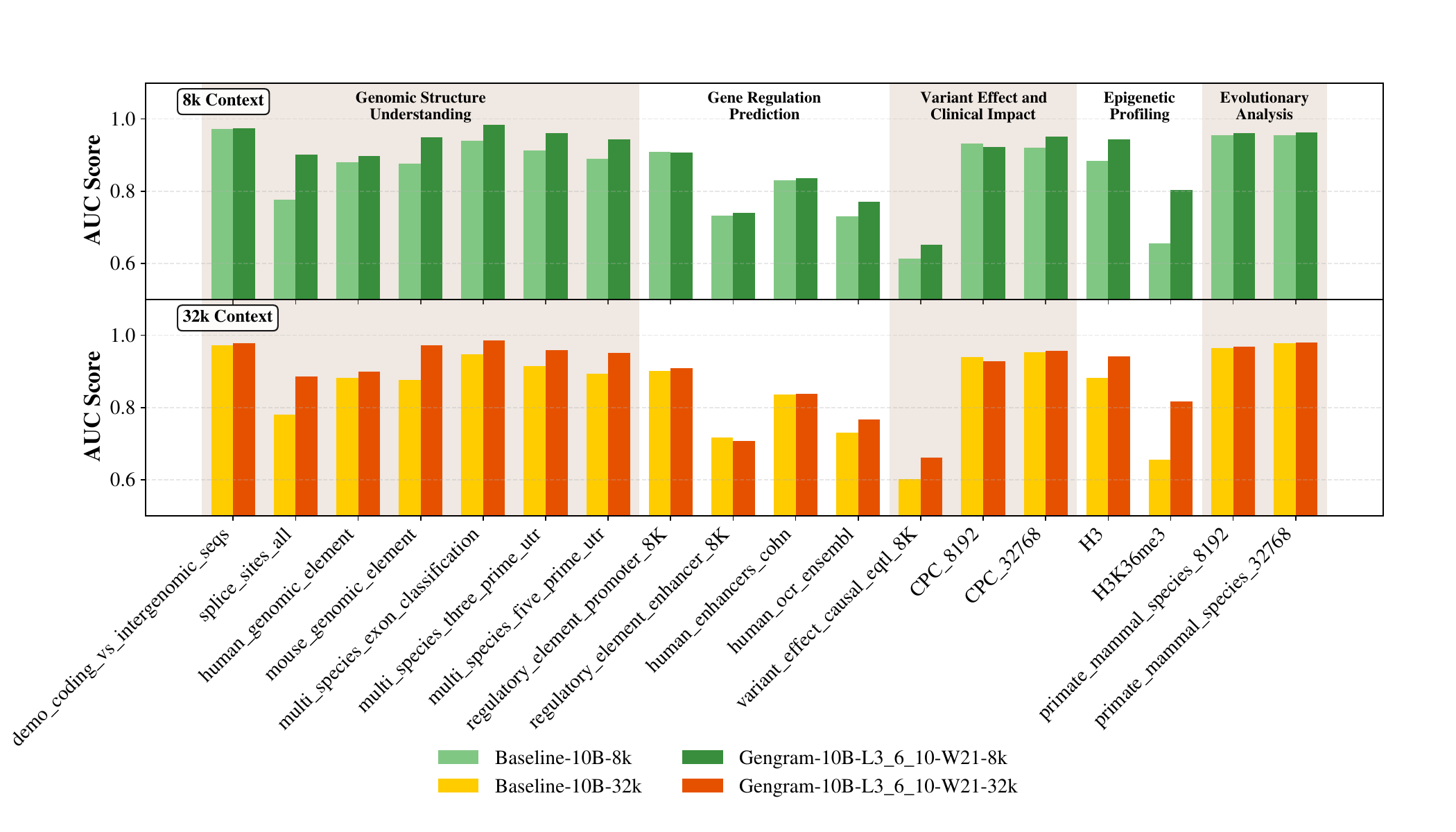}}
    \caption{
      Performance Comparison of Baseline and Gengram Models at 8k and 32k Context Lengths
   }
    \label{fig:benchmark_combine}
  \end{center}
\end{figure*}

A systematic evaluation demonstrates the significant contribution of the Gengram module to genomic sequence modeling. Compared to the Baseline-10B-8k, as can be seen from Figure \ref{fig:benchmark_combine} top, the Gengram-10B-L3\_6\_10-W21-8k model exhibits consistent improvements across almost all assessed metrics. This comprehensive enhancement underscores the overall effectiveness of the Gengram module in strengthening the model's capacity to capture intricate genomic patterns. Furthermore, an ablation study on the window mechanism—comparing Gengram-10B-L3\_6\_10-W21-8k with its window-free counterpart, as illustrated in Table \ref{table:cpt}, Gengram-10B-L3\_6\_10-8k—reveals that the local window contributes substantially to these gains. The performance improvement is particularly pronounced in tasks requiring the recognition of fine-grained local sequence patterns, such as splice site detection and histone mark prediction, highlighting the mechanism's role in processing immediate genomic context.

The Gengram module's capability is further underscored when comparing the Gengram-10B-L3\_6\_10-W21-8k model to the Baseline-10B-32k (see Figure \ref{fig:benchmark_combine}). The former, trained on 8k sequences, achieves performance that not only matches but exceeds the latter, which was trained on significantly longer 32k sequences based on Baseline-10B-8k. This result indicates that the Gengram module confers a powerful long-context modeling ability, effectively compensating for the shorter nominal context length without requiring the substantial additional computational cost typically associated with continued pretraining (CPT) on extended sequences. The most compelling evidence for the window mechanism's intrinsic long-context capability comes from the CPT experiment. Adapting the window-less Gengram-10B-L3\_6\_10-8k to 32k sequences yields a massive performance boost. In stark contrast, applying the same CPT to the window-equipped Gengram-10B-L3\_6\_10-W21-8k model results in only marginal gains (as shown in table \ref{table:cpt}). This stark divergence strongly suggests that the local window mechanism itself inherently equips the model with a capability analogous to long-context modeling.

This phenomenon can be interpreted through the lens of inductive bias. The local window acts as a powerful architectural constraint that guides the model to prioritize locally contiguous information. When this mechanism is stacked across multiple layers of the network, it progressively expands the model's effective receptive field. This structured approach allows the model to efficiently build meaningful representations of extensive genomic regions by systematically composing local features, thereby learning to capture longer-range dependencies even within a shorter nominal context window. This inductive bias not only enhances local pattern recognition but fundamentally strengthens the model's capacity for integrated long-range dependency modeling, providing a computationally efficient pathway to sophisticated genomic sequence analysis.
\begin{table}[h!]
  \caption{CPT Adaptation Results for Models With and Without Local Window.}
  \label{table:cpt}
  \begin{center}
    \begin{small}
      \begin{sc}
        \begin{tabular}{llcccc} 
          \toprule
                                                                                 &                                                                                 & \begin{tabular}[c]{@{}l@{}}Gengram-\\ 10B-\\ L3\_6\_10-8k\end{tabular} & \begin{tabular}[c]{@{}l@{}}Gengram-\\ 10B-\\ L3\_6\_10-\\W21-8k\end{tabular} & \begin{tabular}[c]{@{}l@{}}Gengram-\\ 10B-\\ L3\_6\_10-32k\end{tabular} & \begin{tabular}[c]{@{}l@{}}Gengram-\\ 10B-\\ L3\_6\_10-\\W21-32k\end{tabular} \\

          \midrule
\multirow{7}{*}{\begin{tabular}[c]{@{}l@{}}Genomic \\ Structure \\ Understanding\end{tabular}}    & \begin{tabular}[c]{@{}l@{}}demo\_coding\_vs\_\\ intergenomic\_seqs\end{tabular} & 0.9412                                                                 & 0.9733                                                                     & 0.9818                                                                  & 0.9776                                                                      \\
\addlinespace
                                                                                                  & splice\_sites\_all                                                              & 0.8453                                                                 & 0.9009                                                                     & 0.8465                                                                  & 0.8858                                                                      \\
                                                                                                  \addlinespace
                                                                                                  & \begin{tabular}[c]{@{}l@{}}human\_\\ genomic\_element\end{tabular}              & 0.8842                                                                 & 0.8982                                                                     & 0.8911                                                                  & 0.8998                                                                      \\
                                                                                                  \addlinespace
                                                                                                  & \begin{tabular}[c]{@{}l@{}}mouse\_\\ genomic\_element\end{tabular}              & 0.9603                                                                 & 0.9495                                                                     & 0.9422                                                                  & 0.9722                                                                      \\
                                                                                                  \addlinespace
                                                                                                  & \begin{tabular}[c]{@{}l@{}}multi\_species\_\\ exon\_classification\end{tabular} & 0.8903                                                                 & 0.9832                                                                     & 0.9794                                                                  & 0.9858                                                                      \\
                                                                                                  \addlinespace
                                                                                                  & \begin{tabular}[c]{@{}l@{}}multi\_species\_\\ three\_prime\_utr\end{tabular}    & 0.8759                                                                 & 0.9601                                                                     & 0.9603                                                                  & 0.9598                                                                      \\
                                                                                                  \addlinespace
                                                                                                  & \begin{tabular}[c]{@{}l@{}}multi\_species\_\\ five\_prime\_utr\end{tabular}     & 0.8397                                                                 & 0.9443                                                                     & 0.9521                                                                  & 0.9508                                                                      \\
                                                                                                  \addlinespace
\multirow{4}{*}{\begin{tabular}[c]{@{}l@{}}Gene Regulation \\ Prediction\end{tabular}}            & \begin{tabular}[c]{@{}l@{}}regulatory\_element\_\\ promoter\_8K\end{tabular}    & 0.8861                                                                 & 0.9074                                                                     & 0.9121                                                                  & 0.9097                                                                      \\
\addlinespace
                                                                                                  & \begin{tabular}[c]{@{}l@{}}regulatory\_element\_\\ enhancer\_8K\end{tabular}    & 0.7054                                                                 & 0.7401                                                                     & 0.7267                                                                  & 0.7065                                                                      \\
                                                                                                  \addlinespace
                                                                                                  & human\_enhancers\_cohn                                                          & 0.7605                                                                 & 0.8358                                                                     & 0.8513                                                                  & 0.8375                                                                      \\
                                                                                                  \addlinespace
                                                                                                  & human\_ocr\_ensembl                                                             & 0.7216                                                                 & 0.7714                                                                     & 0.7785                                                                  & 0.7663                                                                      \\
            \midrule
\multirow{3}{*}{\begin{tabular}[c]{@{}l@{}}Variant Effect \\ and \\ Clinical Impact\end{tabular}} & \begin{tabular}[c]{@{}l@{}}variant\_effect\_\\ causal\_eqtl\_8K\end{tabular}    & 0.6078                                                                 & 0.6518                                                                     & 0.6189                                                                  & 0.6607                                                                      \\
\addlinespace
                                                                                                  & CPC\_8192                                                                       & 0.8875                                                                 & 0.9219                                                                     & 0.9269                                                                  & 0.9283                                                                      \\
                                                                                                  \addlinespace
                                                                                                  & CPC\_32768                                                                      & 0.864                                                                  & 0.952                                                                      & 0.9606                                                                  & 0.9572                                                                      \\
            \midrule
\multirow{2}{*}{\begin{tabular}[c]{@{}l@{}}Epigenetic \\ Profiling\end{tabular}}                  & H3                                                                              & 0.8934                                                                 & 0.9434                                                                     & 0.941                                                                   & 0.9417                                                                      \\
\addlinespace
                                                                                                  & H3K36me3                                                                        & 0.7227                                                                 & 0.8039                                                                     & 0.8045                                                                  & 0.8161                                                                      \\
            \midrule
\multirow{2}{*}{\begin{tabular}[c]{@{}l@{}}Evolutionary \\ Analysis\end{tabular}}                 & \begin{tabular}[c]{@{}l@{}}primate\_mammal\_\\ species\_8192\end{tabular}       & 0.7275                                                                 & 0.9611                                                                     & 0.9676                                                                  & 0.9691                                                                      \\
\addlinespace
                                                                                                  & \begin{tabular}[c]{@{}l@{}}primate\_mammal\_\\ species\_32768\end{tabular}      & 0.6933                                                                 & 0.963                                                                      & 0.9785                                                                  & 0.9793                                                                    \\
          \bottomrule
        \end{tabular}
      \end{sc}
    \end{small}
  \end{center}
  \vskip -0.1in
\end{table}

\subsection{Benchmarking Results: State-of-the-Art Performance with Superior Efficiency}
Despite being trained on a significantly smaller dataset of only 200B tokens—an order of magnitude less than competitors like Genos-10B (2.2T) and Evo2-40B (9.3T)  — and despite utilizing significantly fewer activated parameters (2.87B activated parameters in Gengram-10B, compared to 40.3B total/activated parameters in Evo2-40B and 3B in GENERATOR-3B, as shown in Table \ref{atable:various GFMs}), the Gengram-10B-L3\_6\_10-W21-8k model demonstrates highly competitive and often superior performance across key genomic benchmarks (see Table \ref{table:vs_other_models}). It achieves state-of-the-art AUC score in Multi-species Exon Classification (0.9832) and excels in Splice Site Identification (0.9009), as shown in Table \ref{table:vs_other_models}, significantly outperforming its direct counterpart Genos-10B. Furthermore, Gengram-10B shows strong evolutionary analysis capabilities, outperforming the much larger Evo2-40B in Primate-Mammal Species Classification tasks. This combination of top-tier performance in specific tasks like genomic structure understanding and its remarkable training efficiency highlights Gengram-10B as an exceptionally parameter-effective model for genomic sequence analysis.
\begin{table*}[h!]
  \caption{\textbf{Benchmarking Results:} GENGRAM-10B-L3\_6\_10-W21-8k vs other GFMs. The first place is in bold; the second placeis in italics.}
  \label{table:vs_other_models}
  \begin{center}
    \begin{small}
      \begin{sc}
        \begin{tabular}{llccccc} 
          \toprule
                                                    & Model                                           & \makecell[c]{Gengram-\\10B-L3\_6\_10\\-W21-8k} & \makecell[c]{Genos-\\10B}       & \makecell[c]{Evo2-\\40B}        & \makecell[c]{GENERator\\-3B}    & \makecell[c]{NT-v3-\\PRETRAIN\\(650M)}  \\
                                                
          \midrule

                                                    & Trained Tokens                                  & \makecell[c]{\textbf{200B}\\($1\times$)}                         & \makecell[c]{2.2T\\($11\times$)}          & \makecell[c]{9.3T\\($46.5\times$)}          & \makecell[c]{386B\\($1.93\times$)}            & \makecell[c]{11T\\($55\times$)}             \\
            \midrule
\multirow{7}{*}{\makecell[l]{Genomic \\Structure \\ Understanding}}    & \makecell[l]{demo\_coding\_vs\_\\intergenomic\_seqs}            & 0.9733                       & \textbf{0.9914} & \textit{0.9886} & 0.9855          & 0.983           \\
                                                    & splice\_sites\_all                              & \textit{0.9009}              & 0.7990          & \textbf{0.9138} & 0.8071          & 0.8644          \\
                                                    & \makecell[l]{human\_\\genomic\_element}                         & 0.8982                       & 0.8890          & 0.9103          & \textit{0.9173} & \textbf{0.9189} \\
                                                    & \makecell[l]{mouse\_\\genomic\_element}                         & 0.9495                       & \textit{0.9521} & \textbf{0.9888} & 0.9015          & 0.9092          \\
                                                    & \makecell[l]{multi\_species\_\\exon\_classification}            & \textbf{0.9832}              & \textit{0.9755} & 0.9332          & 0.9455          & 0.9615          \\
                                                    & \makecell[l]{multi\_species\_\\three\_prime\_utr}               & \textit{0.9601}              & 0.9511          & \textbf{0.9634} & 0.9400          & 0.9408          \\
                                                    & \makecell[l]{multi\_species\_\\five\_prime\_utr}                & \textit{0.9443}              & 0.9317          & \textbf{0.9895} & 0.8841          & 0.9084          \\
            \midrule
\multirow{4}{*}{\makecell[l]{Gene Regulation \\ Prediction}}         & \makecell[l]{regulatory\_element\_\\promoter\_8K}               & 0.9074                       & \textbf{0.9249} & \textit{0.9227} & 0.9195          & 0.922           \\
                                                    & \makecell[l]{regulatory\_element\_\\enhancer\_8K}               & 0.7401                       & \textbf{0.7532} & \textit{0.7527} & 0.7390          & 0.7415          \\
                                                    & human\_enhancers\_cohn                          & \textit{0.8358}              & \textbf{0.8552} & 0.7756          & 0.8181          & 0.8184          \\
                                                    & human\_ocr\_ensembl                             & \textbf{0.7714}              & 0.7623          & \textit{0.7635} & 0.7270          & 0.7426          \\
            \midrule
\multirow{3}{*}{\makecell[l]{Variant Effect \\ and \\Clinical Impact}} & \makecell[l]{variant\_effect\_\\causal\_eqtl\_8K}               & 0.6518                       & 0.6773          & \textbf{0.7054} & \textit{0.6920} & 0.687           \\
                                                    & CPC\_8192                                       & 0.9219                       & \textbf{0.9522} & 0.9401          & 0.9315          & \textit{0.9468} \\
                                                    & CPC\_32768                                      & 0.952                        & \textbf{0.9625} & \textit{0.9611} & 0.9237          & 0.9426          \\
            \midrule
\multirow{2}{*}{\makecell[c]{Epigenetic \\Profiling}}               & H3                                              & \textit{0.9434}              & 0.9400          & 0.9311          & 0.9163          & \textbf{0.9505} \\
                                                    & H3K36me3                                        & 0.8039                       & 0.7658          & \textbf{0.8823} & 0.8247          & \textit{0.8535} \\
            \midrule
\multirow{2}{*}{\makecell[l]{Evolutionary \\Analysis}}              & \makecell[l]{primate\_mammal\_species\_\\classification\_8192}  & \textit{0.9611}              & \textbf{0.9739} & 0.8867          & 0.8424          & 0.8652          \\
                                                    & \makecell[l]{primate\_mammal\_species\_\\classification\_32768} & \textit{0.9630}               & \textbf{0.9804} & 0.9101          & 0.8709          & 0.8928          \\
          \bottomrule
        \end{tabular}
      \end{sc}
    \end{small}
  \end{center}
  \vskip -0.1in
\end{table*}

\subsection{Generalizability Assessment of Gengram}
\begin{figure}[ht]
  \vskip 0.1in
  \begin{center}
    \centerline{\includegraphics[width=0.5\columnwidth]{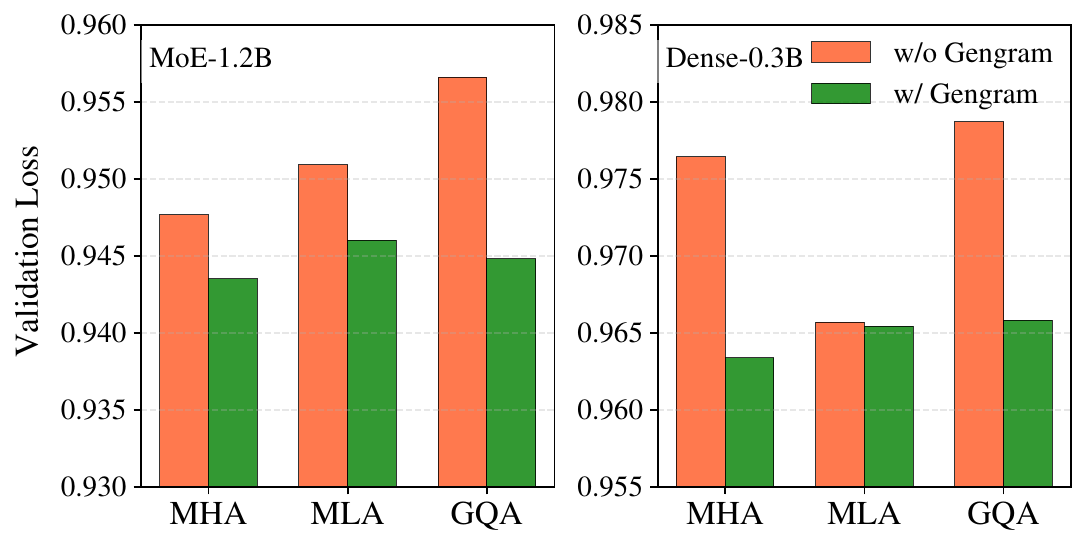}}
    \caption{
      \textbf{Validation-Loss Comparison:}Validation-loss bar chart under MoE and Dense architectures across GQA/MLA/MHA.
    }
    \label{fig:Generalizability}
  \end{center}
\end{figure}

The results are shown in Figure \ref{fig:Generalizability}, where we observe that across different architectures and different attention mechanisms, adding the Gengram module consistently leads to lower validation loss.
Gengram is integrated into each Transformer block as a residual branch, which enables strong compatibility with different network architectures. To systematically verify its cross-architecture generality and consistent gains, we conduct controlled experiments across two architectural families, including MoE (1.2B total parameters, 0.3B activated parameters) and Dense (0.3B parameters) \cite{liu2024deepseek}. Within each family, we further perform cross integration with three mainstream attention mechanisms, Multi-Head Attention(MHA; \cite{vaswani2017attention}), Multi-Head Latent Attention(MLA; \cite{liu2024deepseek}), and Grouped-Query Attention(GQA; \cite{ainslie2023gqa}), and train all settings for 50B tokens. This yields 6 experimental groups for a systematic comparison of Gengram’s adaptability and effectiveness (detailed configurations are provided in Appendix Table \ref{atable:MHA}, Appendix Table \ref{atable:MLA} and Appendix Table \ref{atable:GQA}). These results indicate that Gengram has strong adaptability and stable gains; moreover, as a lightweight module with extremely small parameter size, it can reliably improve performance without increasing training overhead, and has the potential to become an essential component of GFMs in the future.

\subsection{Gengram Stabilizes MoE's Load Balancing}
\begin{figure}[ht]
  \vskip 0.2in
  \begin{center}
    \centerline{\includegraphics[width=0.5\columnwidth]{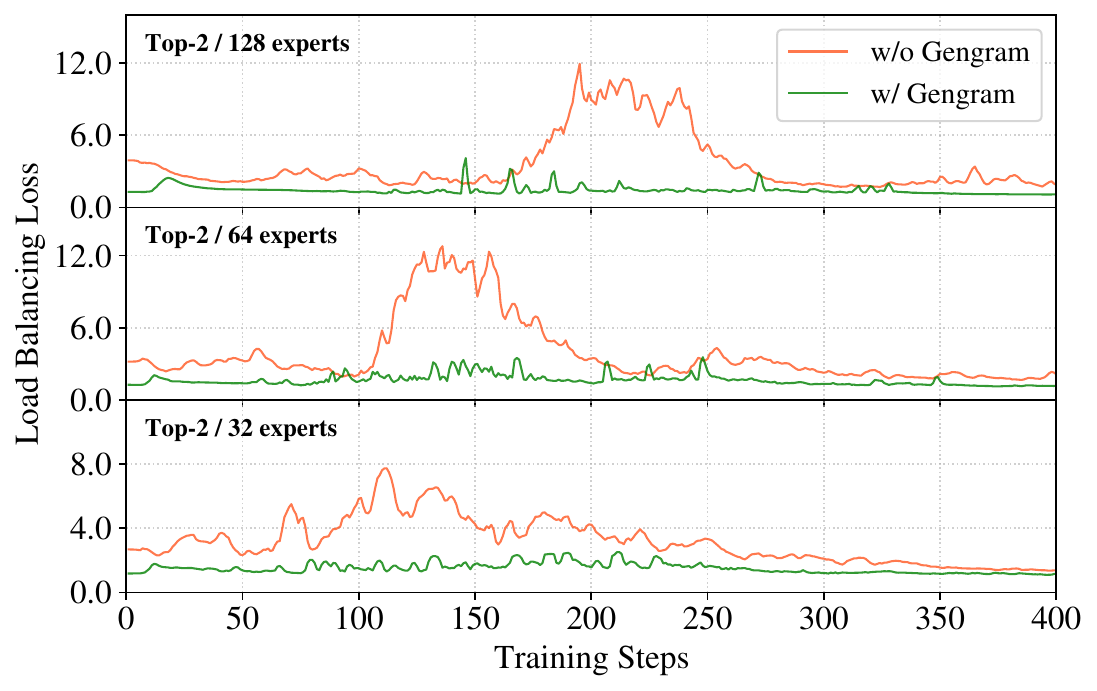}}
    \caption{
       \textbf{Cross-Sparsity Load-Balancing:}Load-balancing loss curves with and without the Gengram module under Top-2 / 128, 64, and 32 experts, showing consistent stabilization across sparsity settings.
    }
    \label{fig:load balance}
  \end{center}
\end{figure}

The load-balancing results in Figure \ref{fig:load balance} show that Gengram exhibits strong stability across different sparsity levels. 

Load imbalance instability is frequently observed in mainstream MoE architectures, and has long been a key challenge during sparse model training; for example, the ST-MoE model introduces router $z$-loss during training to improve stability \cite{zoph2022st}. During Gengram training, we observed that load balancing is more stable than training without Gengram. To validate this observation, we conducted load-balancing tests on a 10B model while keeping all other conditions unchanged, under Top-2 / 128 experts, Top-2 / 64 experts, and Top-2 / 32 experts (Detailed configurations are provided in Appendix Table \ref{atable:10Bmodel structure}), and monitored the load-balancing loss curves.

Under sparse computation, load imbalance often arises from a “positive feedback chain”: routing/activation scores are contaminated by task-irrelevant high-frequency noise, and once such noise is repeatedly injected, it amplifies the advantage of certain routes/keys, eventually causing a small number of paths to be persistently selected \cite{cheng2025ermoe}. Meanwhile, genomic data further exacerbates imbalance: substantial evidence indicates that repetitive and low-complexity regions occupy a large fraction of the genome, and many $k$-mers are highly frequent yet weakly informative for biological function \cite{haubold2006repetitive,li2014diminishing}. We argue that the key of Gengram is that the module first absorbs such local-pattern noise evidence into the memory branch, but does not write it back to the backbone unconditionally: we introduce a gate at the output of memory aggregation; gating can increase sparsity \cite{qiu2025gated} and filter out noise, such that when the backbone is insufficient to distinguish functional motifs from noise, the gate attenuates this memory injection, thereby breaking the feedback loop; consequently, even at higher sparsity, the variance of activation/routing scores is suppressed, and the load distribution becomes closer to uniform and more stable.

\subsection{Intrinsic Mechanisms and Interpretability}

\subsubsection{Window Selection}
We analyze the norm of the block output hidden states: for each position $t$, we take the final hidden state vector after self-attention and Gengram residual injection, and compute its $L2$ norm
\begin{equation}
    \|h_t\|_2 \leftarrow h + \mathrm{Gengram}(h, \text{input\_ids})
    \label{equ17}
\end{equation}
This metric captures the position-wise magnitude of representations and serves as an interpretable signal closely related to numerical stability and the strength of residual updates. Since the hidden state already includes the attention contribution, the observed changes reflect how Gengram modulates the entire block representation via its residual update, rather than the Engram branch alone.
We take a checkpoint trained with the non-windowed Gengram configuration and run two inference-time forward passes:

(i) the original non-windowed implementation to obtain the position-wise $\|h_t\|_2$ curve;

(ii) keeping the checkpoint weights unchanged, we replace the Engram readout/aggregation with a windowed variant (e.g., window size 21) and recompute $\|h_t\|_2$.

\begin{figure}[ht]
  \vskip 0.2in
  \begin{center}
    \centerline{\includegraphics[width=0.5\columnwidth]{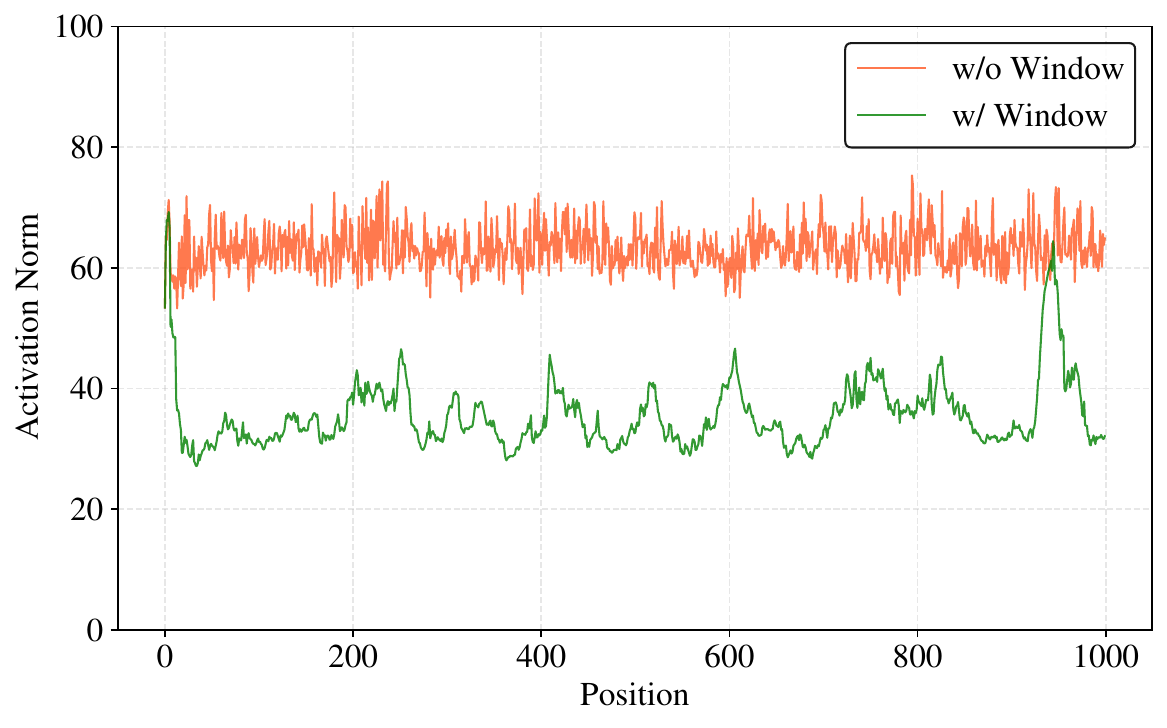}}
    \caption{
      \textbf{Windowed Gengram stabilizes block output representation magnitude.} We plot the position-wise L2 norm of the Transformer block output hidden states $\|h_t\|_2$, which includes self-attention and the Gengram residual injection. Using the same checkpoint trained without windowing, we compare inference-time forward passes with (green) vs. without (orange) windowed Gengram retrieval. Windowing produces smoother and more bounded magnitude dynamics, suggesting that local aggregation in memory readout suppresses spiky residual perturbations and stabilizes the composition between Engram and attention updates.
    }
    \label{fig:Windowed Gengram stabilizes}
  \end{center}
\end{figure}

As shown in Figure \ref{fig:Windowed Gengram stabilizes}, this is a “structure intervention with frozen weights”: if the windowed variant substantially changes the magnitude fluctuation pattern of hidden states, it indicates that the window’s effect is not due to additional training, but due to a structural constraint on memory retrieval/aggregation, which alters the statistics of the residual update and thus the block’s output dynamics.

In the non-windowed implementation, $\|h_t\|_2$ exhibits more pronounced high-frequency jitter and spiky fluctuations across positions, suggesting that the Gengram residual injection introduces more uneven magnitude perturbations along the sequence. After introducing a window (e.g., 21 bp), the $\|h_t\|_2$ curve becomes smoother, local spikes are suppressed, and the magnitude variation is more bounded/regulated. This indicates that windowed retrieval converts highly position-sensitive “point-like” memory triggering into a more consistent local aggregation response, reducing abrupt local changes in representation magnitude—an effect of structural stabilization.

Importantly, since we measure the full block output, this stability implies not only a smoother Engram branch, but also reduced uncertainty in how the Engram residual composes with the attention residual, yielding more stable hidden states overall.

\subsubsection{Mechanistic Interpretation of Training Acceleration}

Gengram accelerates optimization by front-loading predictive computation: shallow representations become aligned with the model’s final decision substantially earlier, effectively reducing the effective depth required to reach a prediction-ready state \cite{cheng2026conditional}.

We quantify layer-wise prediction readiness with a LogitLens-style diagnostic\cite{nostalgebraist2020logitlens}. For each layer hidden state $h^{(l)}$, we apply the final LM head to obtain an intermediate next-token distribution $p^{(l)}(\cdot\mid x)$. We then compute
$\mathrm{KL}\!\left(p^{(L)}(\cdot\mid x)\ \|\ p^{(l)}(\cdot\mid x)\right)$,
where $p^{(L)}$ is the final-layer norm distribution. Smaller KL means that the layer $l$ representation, under a fixed readout, already induces a distribution close to the model’s final decision—i.e., it is closer to being prediction-ready.

Benefiting from Gengram, at injected layers ( $l\in\{3,6,10\}$), the model performs a residual update $h^{(l)} \leftarrow h^{(l)} + \mathrm{engram}\!\left(h^{(l)}, \mathrm{input\_ids}\right)$, which explicitly writes reusable motif memory into the residual stream. This changes the allocation of work across depth: rather than requiring deeper layers to repeatedly reconstruct local predictive primitives from scratch, Gengram makes those primitives available earlier and in a form that is directly consumable by the same final head. The local window constraint ($W{=}21$) further stabilizes this contribution by enforcing local, phase-consistent aggregation, limiting noisy or inconsistent retrieval and enabling intermediate states to settle into a prediction-aligned regime sooner.

Earlier prediction readiness shortens the effective depth over which the model must transform representations before they become decision-relevant. Concretely, when shallow layers already produce head-readable predictive structure, gradient credit assignment no longer needs to traverse as many layers to reinforce useful features. The layer-wise KL collapse therefore offers an interpretable mechanism for training acceleration: Gengram shifts predictive representation formation to earlier layers, reducing the burden on deeper layers and improving optimization efficiency under identical training and readout.

\begin{figure}[ht]
  \vskip 0.2in
  \begin{center}
    \centerline{\includegraphics[width=0.4\columnwidth]{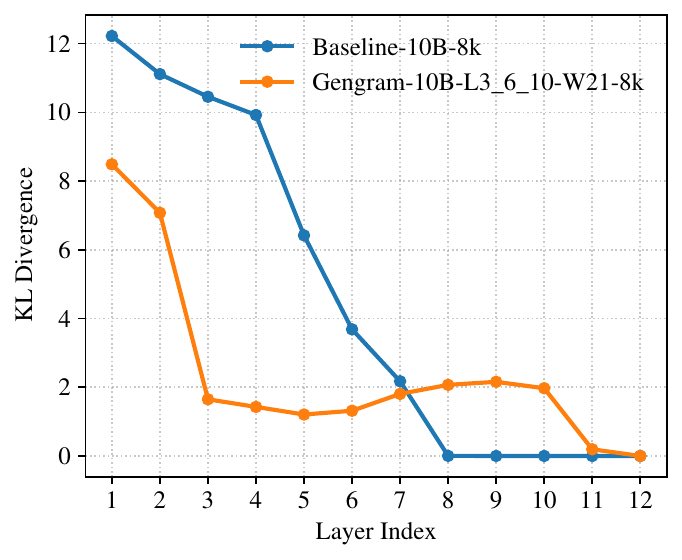}}
    \caption{
      \textbf{Layer-wise prediction readiness via LogitLens-KL:} Compared to baseline-10B-8k, Gengram-10B-L3/6/10-W21-8k exhibits a much faster KL drop in early layers.
    }
    \label{fig:Layer-wise prediction}
  \end{center}
\end{figure}

\subsubsection{Gengram residual contribution aligns with motif}

\begin{figure*}[ht]
  \vskip 0.2in
  \begin{center}
    \centerline{\includegraphics[width=\textwidth]{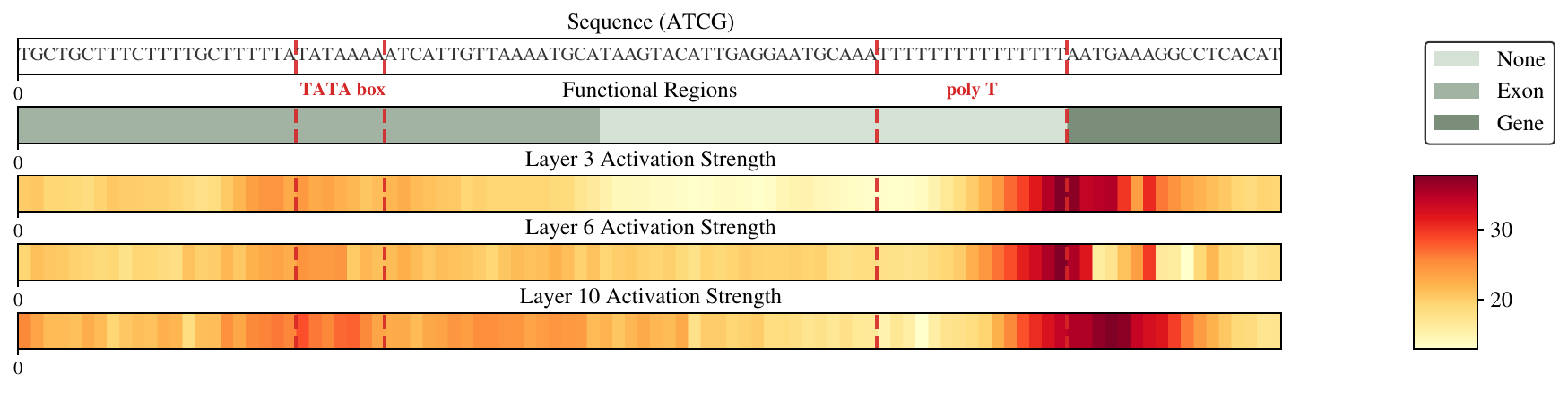}}
    \caption{
      \textbf{Gengram residual contribution aligns with functional elements.} Top: Sequence with motif hits (TATA-like, poly-T) and region labels. Bottom: Gengram residual-write strength $s_\ell(t)$ at L3/6/10. Peaks align with motifs/boundaries and are slightly left-shifted to motif ends, consistent with forward-only windowed matching.
   }
    \label{fig:activation strength}
  \end{center}
\end{figure*}

We find that Gengram’s intervention along a genomic sequence is extremely sparse and high-contrast. Defining the activation strength as $s_\ell(t)=\|\mathbf{r}_\ell(t)\|_2$, where $\qquad
\mathbf{r}_{\ell}(t)=\mathrm{Gengram}_{\ell}(\mathbf{h}_{\ell}(t),x_t)$. Then we plot the per-position residual-write magnitude $s_\ell(t)$, the signal concentrates into sharp peaks occupying only a small fraction of positions, while the majority of the sequence remains near baseline (Fig. \ref{fig:activation strength}). Importantly, these peaks are not randomly distributed: they are enriched around motif-like elements and functional boundaries, including promoter-proximal signals (TATA-like substring) and low-complexity tracts (poly-T).

By design, $s_\ell(t)$ is derived from the additive residual written by Gengram at position $t$. Therefore, a localized peak means that the module chooses to modify the hidden state strongly at that locus. The systematic enrichment near motifs/boundaries supports the hypothesis that Gengram acts as a motif-sensitive memory pathway. While the alignment does not by itself prove causal necessity of the motif for downstream predictions, it establishes a direct mechanistic footprint: Gengram’s writes are targeted to loci carrying concentrated sequence evidence.

The spatial distribution of residual writes changes systematically with depth. In early/mid Gengram layers (L3/L6), the strongest residual writes are tightly anchored to the poly-T or boundary locus: both L3 and L6 reach their maxima at position 82, i.e., at the end of the poly-T tract immediately preceding the intergenic region $\rightarrow$ Gene transition. In contrast, the deepest Engram layer (L10) reaches its maximum at position 86, inside the Gene region, and its top-10 peaks are predominantly in Gene (8/10 for L10 vs. 3/10 for L6). This structured shift—from boundary-adjacent peaks in early/mid layers to gene-internal concentration in deeper layers—supports a hierarchical mechanism: earlier layers act as motif/boundary "writers" that inject localized evidence, while deeper layers integrate and propagate this evidence into more context-dependent representations.

Finally, these patterns are more consistent with motif memory than a rigid motif dictionary. A dictionary-like mechanism would tend to collapse biologically “equivalent” $k$-mers into shared prototypes, producing uniform responses across synonymous realizations. Instead, the observed residual-write signatures are compatible with context-dependent retrieval: biologically related patterns can remain separable when embedded in different surrounding contexts. This is consistent with memory retrieval indexed jointly by substring and hidden state, rather than by a fixed equivalence class over $k$-mers.

\section{Conclusion}
In this work, we introduced Gengram, a conditional motif memory module that addresses a core limitation of Transformer-based GFMs by explicitly encoding and indexing biologically functional multi-nucleotide motifs. Through a gate-controlled memory pathway with local window aggregation, Gengram enables direct motif-level access while preserving the flexibility of standard attention-based sequence modeling. 

Across extensive experiments, Gengram consistently improves both performance and training efficiency over prior state-of-the-art GFMs, yielding substantial gains on motif-dominated tasks, long-range sequence modeling, and improved load balancing behavior. Moreover, Gengram remains effective across diverse architectures and attention mechanisms while introducing negligible computational overhead and enabling stable load balancing in sparse MoE models. 

Beyond empirical improvements, mechanistic analyses reveal that Gengram learns biologically meaningful structure in its memory space, supporting its role as a reusable motif memory rather than a static pattern dictionary. Taken together, these results suggest that Gengram holds promise for scaling GFMs to longer sequence contexts and extending them to multi-omics modeling settings. We believe that Gengram represents an important architectural direction for future GFM development.

\section{Limitations and Future Works}
This work has several limitations that point to important directions for future research. First, our current study primarily validates Gengram as a modular architectural component under relatively constrained training scales. A systematic investigation of scaling behavior—including larger training corpora, longer context lengths, and extended training schedules—remains an important next step. Second, Gengram currently employs fixed k-mer levels and predefined window sizes, motivated by biological priors and computational considerations. Although this design proves effective, exploring adaptive designs on learnable k-mer granularities or adaptive window mechanisms is a promising direction for extending Gengram’s expressiveness. Finally, while our evaluation focuses on standard genomic benchmarks, translating motif-level modeling improvements to application-driven biological settings remains an important challenge. For example, assessing the impact of Gengram on tasks involving subtle genetic variations will require closer integration with downstream biological and experimental validation.

\section*{Impact Statement}
This paper presents work whose goal is to advance the interaction field of Machine
Learning and Genomics. There are many potential societal consequences of our work in this interdisciplinary area, none of which we feel must be specifically highlighted here.

\section*{Acknowledgement}
The model training process was conducted entirely on the 021 Large Science Model and Zero2X open platform. We acknowledge CycloneSEQ for providing high quality sequencing data used in this study. Thanks to the DeepSeek team for their pioneering work in the field of large language models. Their research achievements provided important inspiration for our study. We extend our special thanks to Cheng Wang and Huijun Shen for their significant contributions. Additionally, we appreciate all participants who provided assistance during the research process.

\bibliography{references}
\bibliographystyle{unsrt}  

\newpage
\appendix
\onecolumn


\section{Data Collection and Processing}
\subsection{Data Collection}
The training dataset for this study consists of 145 high-quality haplotype-resolved assemblies, including both human and non-human primate genomes. Human sequences were primarily sourced from the Human Pangenome Reference Consortium (HPRC, release2), supplemented by the GRCh38 and CHM13 reference genomes. Non-human primate sequences were integrated from the NCBI RefSeq database to incorporate evolutionary diversity.
\subsection{Tokenization and Encoding}
All sequences were processed using one-hot encoding. The vocabulary includes the four canonical bases (A, T, C, G), the ambiguous nucleotide N, and the end-of-document token $<$EOD$>$.
\subsection{Two-Stage Data Processing Pipeline}
\subsubsection{Stage I: Base Pre-training (8k Context)}
In the first stage, Datasets 1 and 2 were processed into 8,192 bp fragments \ref{atable:train_dataset_1.2B8k},\ref{atable:train_dataset_10B8k}. To focus on functional representations, we filtered out fragments positioned beyond 5,120 bp of any gene boundary. This yielded a total of 250B tokens, with a 50B subset allocated for architectural ablation studies and the remaining 200B for formal pre-training. Sequences in this stage were restricted to the forward-strand orientation.
\subsubsection{Stage II: Continued Pre-training (CPT, 32k Context)}
In the CPT stage, Dataset 3 was formatted into 32,768 bp fragments \ref{atable:train_dataset_10B32k}. This stage introduced a dual-strand randomized assignment mechanism (56\% negative strand and 44\% positive strand) and removed the gene-distance filtering. Integrating distal intergenic regions and repetitive elements exposes the model to the full genomic landscape, enabling robust feature extraction from functionally 'negative' background sequences.
\begin{table}[h]
  \caption{Datasets for 1.2B model, 50B token total, 8k seq length.}
  \label{atable:train_dataset_1.2B8k}
  \centering 
  \begin{small}
    \begin{sc}
      \begin{tabular}{ll|ll} 
        \toprule
        \textbf{current\_accession} & \textbf{organism\_name} & \textbf{current\_accession} & \textbf{organism\_name}  \\
        \midrule
        Human (HPRC)     & \textit{Homo sapiens} & GCA\_021498455.1 & \textit{Colobus guereza} \\
        GCA\_047047975.1 & \textit{Trachypithecus germaini} & GCA\_026956025.1 & \textit{Macaca cyclopis} \\
        GCA\_049640505.1 & \textit{Gorilla beringei beringei} & GCA\_046862485.1 & \textit{Callithrix penicillata} \\
        GCA\_030222135.1 & \textit{Aotus nancymaae} & GCA\_965153135.1 & \textit{Papio papio} \\
        GCA\_047655295.1 & \textit{Semnopithecus entellus} & GCA\_023783065.1 & \textit{Nomascus siki} \\
        \bottomrule
      \end{tabular}
    \end{sc}
  \end{small}
  \vskip -0.1in
\end{table}

\begin{table}[h!]
  \caption{Datasets for 10B model, 200B token total, 8k seq length.}
  \label{atable:train_dataset_10B8k}
  \centering 
  \begin{small}
    \begin{sc}
      \begin{tabular}{ll|ll} 
        \toprule
        \textbf{current\_accession} & \textbf{organism\_name} & \textbf{current\_accession} & \textbf{organism\_name}  \\
        \midrule
        Human (HPRC)                                     & \textit{Homo sapiens}                       & GCA\_014849445.1                                & \textit{Cercopithecus mona}                 \\
        GCA\_047047975.1                                & \textit{Trachypithecus germaini}            & GCA\_031835075.1                                & \textit{Saguinus oedipus}                   \\
        GCF\_008728515.1                                & \textit{Papio anubis}                       & GCA\_023783475.1                                & \textit{Daubentonia madagascariensis}       \\
        GCF\_009663435.1                                & \textit{Callithrix jacchus}                 & GCF\_009828535.3                                & \textit{Hylobates moloch}                   \\
        GCA\_030128845.1                                & \textit{Pan troglodytes}                    & GCA\_965153135.1                                & \textit{Papio papio}                        \\
        GCA\_000235385.1                                & \textit{Saimiri boliviensis boliviensis}    & GCF\_008122165.1                                & \textit{Gorilla gorilla gorilla}            \\
        GCA\_947095605.1                                & \textit{Pongo pygmaeus}                     & GCA\_026956025.1                                & \textit{Macaca cyclopis}                    \\
        GCA\_046862485.1                                & \textit{Callithrix penicillata}             & GCA\_049640505.1                                & \textit{Gorilla beringei beringei}          \\
        GCF\_003339765.1                                & \textit{Macaca mulatta}                     & GCA\_023764695.1                                & \textit{Pygathrix nigripes}                 \\
        GCA\_008086735.1                                & \textit{Cheirogaleus medius}                & GCA\_048565385.1                                & \textit{Saimiri boliviensis}                \\
        GCA\_028878085.3                                & \textit{Symphalangus syndactylus}           & GCF\_041146395.1                                & \textit{Eulemur rufifrons}                  \\
        GCF\_024542745.1                                & \textit{Macaca thibetana thibetana}         & GCA\_023807365.1                                & \textit{Macaca silenus}                     \\
        GCF\_009764315.1                                & \textit{Trachypithecus francoisi}           & GCA\_047655295.1                                & \textit{Semnopithecus entellus}             \\
        GCA\_023783065.1                                & \textit{Nomascus siki}                      & GCF\_000165445.2                                & \textit{Microcebus murinus}                 \\
        GCA\_040869165.1                                & \textit{Theropithecus gelada}               & GCA\_030222135.1                                & \textit{Aotus nancymaae}                    \\
        GCA\_023783095.1                                & \textit{Macaca assamensis}                  & GCA\_040437455.1                                & \textit{Plecturocebus cupreus}              \\
        GCA\_047496115.1                                & \textit{Brachyteles arachnoides}            & GCA\_023783575.1                                & \textit{Cebus albifrons}                    \\
        GCF\_012559485.2                                & \textit{Macaca fascicularis}                & GCA\_047834645.1                                & \textit{Nasalis larvatus}                   \\
        GCA\_021498455.1                                & \textit{Colobus guereza}                    &              \\
        \bottomrule
      \end{tabular}
    \end{sc}
  \end{small}
  \vskip -0.1in
\end{table}

\begin{table}[h]
  \caption{Datasets for 10B model, 100B token total, 32k seq length..}
  \label{atable:train_dataset_10B32k}
  \centering 
  \begin{small}
    \begin{sc}
      \begin{tabular}{ll|ll} 
        \toprule
        \textbf{current\_accession} & \textbf{organism\_name} & \textbf{current\_accession} & \textbf{organism\_name}  \\
        \midrule
        Human(HPRC)                                     & \textit{Homo sapiens}                       & GCA\_003339765.3                                & \textit{Macaca mulatta}                     \\
        GRCh38                                          & \textit{Homo sapiens}                       & GCA\_013052645.3                                & \textit{Pan paniscus}                       \\
        GCA\_002880755.3                                & \textit{Pan troglodytes}                    & GCA\_023764695.1                                & \textit{Pygathrix nigripes}                 \\
        GCF\_028878055.3                                & \textit{Symphalangus syndactylus}           & GCA\_007565055.1                                & \textit{Rhinopithecus roxellana}            \\
        GCF\_030222135.1                                & \textit{Aotus nancymaae}                    & GCA\_023767775.1                                & \textit{Pongo pygmaeus}                     \\
        GCA\_028878055.3                                & \textit{Symphalangus syndactylus}           & GCA\_965153045.1                                & \textit{Macaca nemestrina}                  \\
        GCF\_003255815.1                                & \textit{Theropithecus gelada}               & GCF\_041146395.1                                & \textit{Lemur catta}                        \\
        GCA\_021498475.1                                & \textit{Saguinus midas}                     & GCA\_023761135.1                                & \textit{Symphalangus syndactylus}           \\
        GCA\_903645265.1                                & \textit{Macaca fascicularis}                & GCA\_012559485.3 & \textit{Macaca fascicularis} \\
        \bottomrule
      \end{tabular}
    \end{sc}
  \end{small}
  \vskip -0.1in
\end{table}

\begin{table}[h!]
  \caption{Benchmarks Datasets.}
  \label{atable:Benchmarks Datases}
  \centering 
  \begin{small}
    \begin{sc}
      \begin{tabular}{llccc} 
        \toprule
                                                            & Task                                            & Source & Length (bp) & Classifier \\

        \midrule
        \multirow{7}{*}{\begin{tabular}[l]{@{}l@{}}Genomic \\ Structure\\  Understanding\end{tabular}} & \begin{tabular}[l]{@{}l@{}}demo\_coding\_vs\\ \_intergenomic\_seqs\end{tabular}            & GB     & 200         & MLP        \\
        \addlinespace
                                                                                               & splice\_sites\_all                                                                         & NTB    & 400         & MLP        \\
                                                                                               \addlinespace
                                                                                               & human\_genomic\_element                                                                    & GeB    & up to 1024  & MLP        \\
                                                                                               \addlinespace
                                                                                               & mouse\_genomic\_element                                                                    & GeB    & up to 1024  & MLP        \\
                                                                                               \addlinespace
                                                                                               & \begin{tabular}[l]{@{}l@{}}multi\_species\_\\ exon\_classification\end{tabular}            & GeB    & 1024        & MLP        \\
                                                                                               \addlinespace
                                                                                               & \begin{tabular}[l]{@{}l@{}}multi\_species\_\\ three\_prime\_utr\end{tabular}               & GeB    & 1024        & MLP        \\
                                                                                               \addlinespace
                                                                                               & \begin{tabular}[l]{@{}l@{}}multi\_species\_\\ five\_prime\_utr\end{tabular}                & GeB    & 1024        & MLP        \\
        \midrule
    \multirow{4}{*}{\begin{tabular}[l]{@{}l@{}}Gene \\ Regulation \\ Prediction\end{tabular}}      & \begin{tabular}[l]{@{}l@{}}regulatory\_element\_\\ promoter\_8K\end{tabular}               & LRB    & 8192        & MLP        \\
    \addlinespace
                                                                                               & \begin{tabular}[l]{@{}l@{}}regulatory\_element\_\\ enhancer\_8K\end{tabular}               & LRB    & 8192        & MLP        \\
                                                                                               \addlinespace
                                                                                               & human\_enhancers\_cohn                                                                     & GB     & 500         & MLP        \\
                                                                                               \addlinespace
                                                                                               & human\_ocr\_ensembl                                                                        & GB     & up to 593   & MLP        \\
        \midrule
    \multirow{3}{*}{\begin{tabular}[l]{@{}l@{}}Variant Effect and \\ Clinical Impact\end{tabular}} & \begin{tabular}[l]{@{}l@{}}variant\_effect\_\\ causal\_eqtl\_8K\end{tabular}               & LRB    & 8192        & MLP        \\
    \addlinespace
                                                                                               & CPC\_8192                                                                                  & GeB    & 8192        & MLP        \\
                                                                                               \addlinespace
                                                                                               & CPC\_32768                                                                                 & GeB    & 32768       & MLP        \\
        \midrule
    \multirow{2}{*}{Epigenetic Profiling}                                                          & H3                                                                                         & NTB    & up to 500   & MLP        \\
    \addlinespace
                                                                                               & H3K36me3                                                                                   & NTB    & up to 500   & MLP        \\
        \midrule
        \multirow{2}{*}{Evolutionary Analysis}                                                         & \begin{tabular}[l]{@{}l@{}}primate\_mammal\_species\_\\ classification\_8192\end{tabular}  & GeB    & 8192        & XGBoost    \\
        \addlinespace
                                                                                               & \begin{tabular}[l]{@{}l@{}}primate\_mammal\_species\_\\ classification\_32768\end{tabular} & GeB    & 32768       & XGBoost \\
        \bottomrule
      \end{tabular}
    \end{sc}
  \end{small}
  \vskip -0.1in
\end{table}

\begin{table}[h!]
  \caption{Comparison of various GFMs.}
  \label{atable:various GFMs}
  \centering
    \begin{small}
      \begin{sc}
        \begin{tabular}{lcccccc} 
          \toprule
          Model & \makecell[c]{Total \\ Parameters} & \makecell[c]{Activated\\ Parameters} & \makecell[c]{Trainned \\ Tokens} & \makecell[c]{Seq. \\Length} & Tokenizer & Dataset \\
          \midrule

            \makecell[l]{Gengram-\\10B-\\L3\_6\_10-\\W21-8k} & 10B              & 2.87B             & 200B           & 8K       & Single Base & \makecell[c]{HPRC,\\Refseq Database}         \\
             \addlinespace
            Genos-10B                 & 10B              & 2.87B             & 2.2T           & 1M        & Single Base & \makecell[c]{HPRC,HGSVC,\\CEPH,GRCH38,\\CHM13} \\
         
          \addlinespace
            Evo2-40B                  & 40.3B            & 40.3B             & 9.3T           & 1M        & Single Base & OpenGenome2                  \\
          \addlinespace
            \makecell[l]{NTv3-\\PRETRAIN}             & 650M             & 650M              & 11T            & 1M        & Single Base & OpenGenome2                  \\
          \addlinespace
            \makecell[l]{GENERator-\\3B}              & 3B               & 3B                & 386B           & 98k       & 6-mer       & \makecell[c]{Refseq eukaryotic \\genomes}    \\ 

          \bottomrule
        \end{tabular}
      \end{sc}
    \end{small}
  \vskip -0.1in
\end{table}

\begin{table}[h!]
\caption{MHA-based Gengram Configurations (Dense vs. MoE)}
  \label{atable:MHA}
  \centering 
  \begin{small}
    \begin{sc}
    \begin{tabular}{|l|cc|}
    \hline
    \rowcolor[HTML]{EFEFEF} 
    {\color[HTML]{1F1F1F} Parameter}                   & \multicolumn{1}{c|}{\cellcolor[HTML]{EFEFEF}{\color[HTML]{1F1F1F} 0.3B Dense (Base/Engram)}} & {\color[HTML]{1F1F1F} 1.2B MoE (Base/Engram)} \\ \hline
    {\color[HTML]{1F1F1F} Total Parameters}            & \multicolumn{1}{c|}{{\color[HTML]{1F1F1F} 0.3B}}                                             & {\color[HTML]{1F1F1F} 1.2B}                   \\ \hline
    {\color[HTML]{1F1F1F} Activated Parameters}           & \multicolumn{2}{c|}{{\color[HTML]{1F1F1F} 0.3B}}                                                                                             \\ \hline
    {\color[HTML]{1F1F1F} Hidden Size}                 & \multicolumn{1}{c|}{{\color[HTML]{1F1F1F} 1280}}                                             & {\color[HTML]{1F1F1F} 1024}                   \\ \hline
    {\color[HTML]{1F1F1F} Num Layers}                  & \multicolumn{2}{c|}{{\color[HTML]{1F1F1F} 12}}                                                                                               \\ \hline
    {\color[HTML]{1F1F1F} Attention Heads}             & \multicolumn{2}{c|}{{\color[HTML]{1F1F1F} 16}}                                                                                               \\ \hline
    {\color[HTML]{1F1F1F} Normalization}               & \multicolumn{2}{c|}{{\color[HTML]{1F1F1F} RMSNorm}}                                                                                          \\ \hline
    {\color[HTML]{1F1F1F} Activation}                  & \multicolumn{2}{c|}{{\color[HTML]{1F1F1F} SwiGLU}}                                                                                           \\ \hline
    {\color[HTML]{1F1F1F} Positional Emb}              & \multicolumn{2}{c|}{{\color[HTML]{1F1F1F} RoPE}}                                                                                             \\ \hline
    {\color[HTML]{1F1F1F} Num Experts}                 & \multicolumn{1}{c|}{{\color[HTML]{1F1F1F} \textbackslash{}}}                                 & {\color[HTML]{1F1F1F} 8}                      \\ \hline
    {\color[HTML]{1F1F1F} Router Top-k}                & \multicolumn{1}{c|}{{\color[HTML]{1F1F1F} \textbackslash{}}}                                 & {\color[HTML]{1F1F1F} 2}                      \\ \hline
    {\color[HTML]{1F1F1F} MoE Operators}               & \multicolumn{1}{c|}{{\color[HTML]{1F1F1F} \textbackslash{}}}                                 & {\color[HTML]{1F1F1F} Grouped GEMM}           \\ \hline
    {\color[HTML]{1F1F1F} Load Balancing}              & \multicolumn{1}{c|}{{\color[HTML]{1F1F1F} \textbackslash{}}}                                 & {\color[HTML]{1F1F1F} Aux Loss}               \\ \hline
    {\color[HTML]{1F1F1F} Gengram Layers}              & \multicolumn{2}{c|}{{\color[HTML]{1F1F1F} Optional \{3, 6, 10\}}}                                                                            \\ \hline
    {\color[HTML]{1F1F1F} Gengram Ngram-sizes}         & \multicolumn{2}{c|}{{\color[HTML]{1F1F1F} Optional (1 to 6)}}                                                                                \\ \hline
    {\color[HTML]{1F1F1F} Gengram Window-sizes}        & \multicolumn{2}{c|}{{\color[HTML]{1F1F1F} Optional (21)}}                                                                                    \\ \hline
    {\color[HTML]{1F1F1F} Gengram Conv-kernel-size}    & \multicolumn{2}{c|}{{\color[HTML]{1F1F1F} Optional (4)}}                                                                                     \\ \hline
    {\color[HTML]{1F1F1F} Gengram Embed-dim-per-ngram} & \multicolumn{2}{c|}{{\color[HTML]{1F1F1F} Optional (1024)}}                                                                                  \\ \hline
    {\color[HTML]{1F1F1F} Seq Length}                  & \multicolumn{2}{c|}{{\color[HTML]{1F1F1F} 8192}}                                                                                             \\ \hline
    {\color[HTML]{1F1F1F} Global Batch Size}           & \multicolumn{2}{c|}{{\color[HTML]{1F1F1F} 1024}}                                                                                             \\ \hline
    {\color[HTML]{1F1F1F} Learning Rate}               & \multicolumn{2}{c|}{{\color[HTML]{1F1F1F} 1e-4 to 1e-5}}                                                                                     \\ \hline
    {\color[HTML]{1F1F1F} LR Scheduler}                & \multicolumn{2}{c|}{{\color[HTML]{1F1F1F} Cosine Decay}}                                                                                     \\ \hline
    \end{tabular}

    \end{sc}
  \end{small}
  \vskip -0.1in
\end{table}

\begin{table}[h!]
\caption{MLA-based Gengram Configurations (Dense vs. MoE)}
  \label{atable:MLA}
  \centering 
  \begin{small}
    \begin{sc}
    \begin{tabular}{|l|cc|}
    \hline
    \rowcolor[HTML]{EFEFEF} 
        {\color[HTML]{1F1F1F} Parameter}                   & \multicolumn{1}{c|}{\cellcolor[HTML]{EFEFEF}{\color[HTML]{1F1F1F} 0.3B Dense (Base/Engram)}} & {\color[HTML]{1F1F1F} 1.2B MoE (Base/Engram)} \\ \hline
        {\color[HTML]{1F1F1F} Total Parameters}            & \multicolumn{1}{c|}{{\color[HTML]{1F1F1F} 0.3B}}                                             & {\color[HTML]{1F1F1F} 1.2B}                   \\ \hline
        {\color[HTML]{1F1F1F} Activated Parameters}           & \multicolumn{2}{c|}{{\color[HTML]{1F1F1F} 0.3B}}                                                                                             \\ \hline
        {\color[HTML]{1F1F1F} Hidden Size}                 & \multicolumn{1}{c|}{{\color[HTML]{1F1F1F} 1280}}                                             & {\color[HTML]{1F1F1F} 1024}                   \\ \hline
        {\color[HTML]{1F1F1F} Num Layers}                  & \multicolumn{2}{c|}{{\color[HTML]{1F1F1F} 12}}                                                                                               \\ \hline
        {\color[HTML]{1F1F1F} Attention Heads}             & \multicolumn{2}{c|}{{\color[HTML]{1F1F1F} 16}}                                                                                               \\ \hline
        {\color[HTML]{1F1F1F} KV-LoRA Rank}                & \multicolumn{1}{c|}{{\color[HTML]{1F1F1F} 512}}                                              & {\color[HTML]{1F1F1F} 256}                    \\ \hline
        {\color[HTML]{1F1F1F} QK Head Dim}                 & \multicolumn{1}{c|}{{\color[HTML]{1F1F1F} 128}}                                              & {\color[HTML]{1F1F1F} 64}                     \\ \hline
        {\color[HTML]{1F1F1F} QK Pos-Emb Dim}              & \multicolumn{1}{c|}{{\color[HTML]{1F1F1F} 128}}                                              & {\color[HTML]{1F1F1F} 64}                     \\ \hline
        {\color[HTML]{1F1F1F} V Head Dim}                  & \multicolumn{1}{c|}{{\color[HTML]{1F1F1F} 256}}                                              & {\color[HTML]{1F1F1F} 128}                    \\ \hline
        {\color[HTML]{1F1F1F} Normalization}               & \multicolumn{2}{c|}{{\color[HTML]{1F1F1F} RMSNorm}}                                                                                          \\ \hline
        {\color[HTML]{1F1F1F} Activation}                  & \multicolumn{2}{c|}{{\color[HTML]{1F1F1F} SwiGLU}}                                                                                           \\ \hline
        {\color[HTML]{1F1F1F} Positional Emb}              & \multicolumn{2}{c|}{{\color[HTML]{1F1F1F} RoPE}}                                                                                             \\ \hline
        {\color[HTML]{1F1F1F} Num Experts}                 & \multicolumn{1}{c|}{{\color[HTML]{1F1F1F} \textbackslash{}}}                                 & {\color[HTML]{1F1F1F} 8}                      \\ \hline
        {\color[HTML]{1F1F1F} Router Top-k}                & \multicolumn{1}{c|}{{\color[HTML]{1F1F1F} \textbackslash{}}}                                 & {\color[HTML]{1F1F1F} 2}                      \\ \hline
        {\color[HTML]{1F1F1F} MoE Operators}               & \multicolumn{1}{c|}{{\color[HTML]{1F1F1F} \textbackslash{}}}                                 & {\color[HTML]{1F1F1F} Grouped GEMM}           \\ \hline
        {\color[HTML]{1F1F1F} Load Balancing}              & \multicolumn{1}{c|}{{\color[HTML]{1F1F1F} \textbackslash{}}}                                 & {\color[HTML]{1F1F1F} Aux Loss}               \\ \hline
        {\color[HTML]{1F1F1F} Gengram Layers}              & \multicolumn{2}{c|}{{\color[HTML]{1F1F1F} Optional \{3, 6, 10\}}}                                                                            \\ \hline
        {\color[HTML]{1F1F1F} Gengram Ngram-sizes}         & \multicolumn{2}{c|}{{\color[HTML]{1F1F1F} Optional (1 to 6)}}                                                                                \\ \hline
        {\color[HTML]{1F1F1F} Gengram Window-sizes}        & \multicolumn{2}{c|}{{\color[HTML]{1F1F1F} Optional (21)}}                                                                                    \\ \hline
        {\color[HTML]{1F1F1F} Gengram Conv-kernel-size}    & \multicolumn{2}{c|}{{\color[HTML]{1F1F1F} Optional (4)}}                                                                                     \\ \hline
        {\color[HTML]{1F1F1F} Gengram Embed-dim-per-ngram} & \multicolumn{2}{c|}{{\color[HTML]{1F1F1F} Optional (1024)}}                                                                                  \\ \hline
        {\color[HTML]{1F1F1F} Seq Length}                  & \multicolumn{2}{c|}{{\color[HTML]{1F1F1F} 8192}}                                                                                             \\ \hline
        {\color[HTML]{1F1F1F} Global Batch Size}           & \multicolumn{2}{c|}{{\color[HTML]{1F1F1F} 1024}}                                                                                             \\ \hline
        {\color[HTML]{1F1F1F} Learning Rate}               & \multicolumn{2}{c|}{{\color[HTML]{1F1F1F} 1e-4 to 1e-5}}                                                                                     \\ \hline
        {\color[HTML]{1F1F1F} LR Scheduler}                & \multicolumn{2}{c|}{{\color[HTML]{1F1F1F} Cosine Decay}}                                                                                     \\ \hline
    \end{tabular}

    \end{sc}
  \end{small}
  \vskip -0.1in
\end{table}

\begin{table}[h!]
\caption{GQA-based Gengram Configurations (Dense vs. MoE)}
  \label{atable:GQA}
  \centering 
  \begin{small}
    \begin{sc}
    \begin{tabular}{|l|cc|}
    \hline
    \rowcolor[HTML]{EFEFEF} 
    {\color[HTML]{1F1F1F} Parameter}                   & \multicolumn{1}{c|}{\cellcolor[HTML]{EFEFEF}{\color[HTML]{1F1F1F} 0.3B Dense (Base/Engram)}} & {\color[HTML]{1F1F1F} 1.2B MoE (Base/Engram)} \\ \hline
    {\color[HTML]{1F1F1F} Total Parameters}            & \multicolumn{1}{c|}{{\color[HTML]{1F1F1F} 0.3B}}                                             & {\color[HTML]{1F1F1F} 1.2B}                   \\ \hline
    {\color[HTML]{1F1F1F} Activated Parameters}           & \multicolumn{2}{c|}{{\color[HTML]{1F1F1F} 0.3B}}                                                                                             \\ \hline
    {\color[HTML]{1F1F1F} Hidden Size}                 & \multicolumn{1}{c|}{{\color[HTML]{1F1F1F} 1280}}                                             & {\color[HTML]{1F1F1F} 1024}                   \\ \hline
    {\color[HTML]{1F1F1F} Num Layers}                  & \multicolumn{2}{c|}{{\color[HTML]{1F1F1F} 12}}                                                                                               \\ \hline
    {\color[HTML]{1F1F1F} Attention Heads}             & \multicolumn{2}{c|}{{\color[HTML]{1F1F1F} 16}}                                                                                               \\ \hline
    {\color[HTML]{1F1F1F} GQA Query Groups}            & \multicolumn{2}{c|}{{\color[HTML]{1F1F1F} 8}}                                                                                                \\ \hline
    {\color[HTML]{1F1F1F} Normalization}               & \multicolumn{2}{c|}{{\color[HTML]{1F1F1F} RMSNorm}}                                                                                          \\ \hline
    {\color[HTML]{1F1F1F} Activation}                  & \multicolumn{2}{c|}{{\color[HTML]{1F1F1F} SwiGLU}}                                                                                           \\ \hline
    {\color[HTML]{1F1F1F} Positional Emb}              & \multicolumn{2}{c|}{{\color[HTML]{1F1F1F} RoPE}}                                                                                             \\ \hline
    {\color[HTML]{1F1F1F} Num Experts}                 & \multicolumn{1}{c|}{{\color[HTML]{1F1F1F} \textbackslash{}}}                                 & {\color[HTML]{1F1F1F} 8}                      \\ \hline
    {\color[HTML]{1F1F1F} Router Top-k}                & \multicolumn{1}{c|}{{\color[HTML]{1F1F1F} \textbackslash{}}}                                 & {\color[HTML]{1F1F1F} 2}                      \\ \hline
    {\color[HTML]{1F1F1F} MoE Operators}               & \multicolumn{1}{c|}{{\color[HTML]{1F1F1F} \textbackslash{}}}                                 & {\color[HTML]{1F1F1F} Grouped GEMM}           \\ \hline
    {\color[HTML]{1F1F1F} Load Balancing}              & \multicolumn{1}{c|}{{\color[HTML]{1F1F1F} \textbackslash{}}}                                 & {\color[HTML]{1F1F1F} Aux Loss}               \\ \hline
    {\color[HTML]{1F1F1F} Gengram Layers}              & \multicolumn{2}{c|}{{\color[HTML]{1F1F1F} Optional \{3, 6, 10\}}}                                                                            \\ \hline
    {\color[HTML]{1F1F1F} Gengram Ngram-sizes}         & \multicolumn{2}{c|}{{\color[HTML]{1F1F1F} Optional (1 to 6)}}                                                                                \\ \hline
    {\color[HTML]{1F1F1F} Gengram Window-sizes}        & \multicolumn{2}{c|}{{\color[HTML]{1F1F1F} Optional (21)}}                                                                                    \\ \hline
    {\color[HTML]{1F1F1F} Gengram Conv-kernel-size}    & \multicolumn{2}{c|}{{\color[HTML]{1F1F1F} Optional (4)}}                                                                                     \\ \hline
    {\color[HTML]{1F1F1F} Gengram Embed-dim-per-ngram} & \multicolumn{2}{c|}{{\color[HTML]{1F1F1F} Optional (1024)}}                                                                                  \\ \hline
    {\color[HTML]{1F1F1F} Seq Length}                  & \multicolumn{2}{c|}{{\color[HTML]{1F1F1F} 8192}}                                                                                             \\ \hline
    {\color[HTML]{1F1F1F} Global Batch Size}           & \multicolumn{2}{c|}{{\color[HTML]{1F1F1F} 1024}}                                                                                             \\ \hline
    {\color[HTML]{1F1F1F} Learning Rate}               & \multicolumn{2}{c|}{{\color[HTML]{1F1F1F} 1e-4 to 1e-5}}                                                                                     \\ \hline
    {\color[HTML]{1F1F1F} LR Scheduler}                & \multicolumn{2}{c|}{{\color[HTML]{1F1F1F} Cosine Decay}} \\ \hline
    \end{tabular}

    \end{sc}
  \end{small}
  \vskip -0.1in
\end{table}

\begin{table}[h!]
\caption{10B model with and without Gengram}
  \label{atable:w-wo Gengram}
  \centering 
  \begin{small}
  \begin{sc}
\begin{tabular}{|l|ccc|}
\hline
\rowcolor[HTML]{EFEFEF} 
{\color[HTML]{1F1F1F} Parameter}                   & \multicolumn{1}{c|}{\cellcolor[HTML]{EFEFEF}{\color[HTML]{1F1F1F} Baseline-10B}} & \multicolumn{1}{c|}{\cellcolor[HTML]{EFEFEF}{\color[HTML]{1F1F1F} Gengram-10B-L3\_6\_10}} & {\color[HTML]{1F1F1F} Gengram-10B-L3\_6\_10-W21} \\ \hline
{\color[HTML]{1F1F1F} Total Parameters}            & \multicolumn{3}{c|}{{\color[HTML]{1F1F1F} 10B}}                                                                                                                                                                                 \\ \hline
{\color[HTML]{1F1F1F} Activated Parameters}           & \multicolumn{3}{c|}{{\color[HTML]{1F1F1F} 2.87B}}                                                                                                                                                                               \\ \hline
{\color[HTML]{1F1F1F} Hidden Size}                 & \multicolumn{3}{c|}{{\color[HTML]{1F1F1F} 4096}}                                                                                                                                                                                \\ \hline
{\color[HTML]{1F1F1F} Num Layers}                  & \multicolumn{3}{c|}{{\color[HTML]{1F1F1F} 12}}                                                                                                                                                                                  \\ \hline
{\color[HTML]{1F1F1F} Attention Heads}             & \multicolumn{3}{c|}{{\color[HTML]{1F1F1F} 16}}                                                                                                                                                                                  \\ \hline
{\color[HTML]{1F1F1F} GQA Query Groups}            & \multicolumn{3}{c|}{{\color[HTML]{1F1F1F} 8}}                                                                                                                                                                                   \\ \hline
{\color[HTML]{1F1F1F} Normalization}               & \multicolumn{3}{c|}{{\color[HTML]{1F1F1F} RMSNorm}}                                                                                                                                                                             \\ \hline
{\color[HTML]{1F1F1F} Activation}                  & \multicolumn{3}{c|}{{\color[HTML]{1F1F1F} SwiGLU}}                                                                                                                                                                              \\ \hline
{\color[HTML]{1F1F1F} Positional Emb}              & \multicolumn{3}{c|}{{\color[HTML]{1F1F1F} RoPE}}                                                                                                                                                                                \\ \hline
{\color[HTML]{1F1F1F} Num Experts}                 & \multicolumn{3}{c|}{{\color[HTML]{1F1F1F} 8}}                                                                                                                                                                                   \\ \hline
{\color[HTML]{1F1F1F} Router Top-k}                & \multicolumn{3}{c|}{{\color[HTML]{1F1F1F} 2}}                                                                                                                                                                                   \\ \hline
{\color[HTML]{1F1F1F} MoE Operators}               & \multicolumn{3}{c|}{{\color[HTML]{1F1F1F} Grouped GEMM}}                                                                                                                                                                        \\ \hline
{\color[HTML]{1F1F1F} Load Balancing}              & \multicolumn{3}{c|}{{\color[HTML]{1F1F1F} Aux Loss}}                                                                                                                                                                            \\ \hline
{\color[HTML]{1F1F1F} Gengram Layers}              & \multicolumn{1}{c|}{{\color[HTML]{1F1F1F} \textbackslash{}}}                     & \multicolumn{1}{c|}{{\color[HTML]{1F1F1F} 3, 6, 10}}                                      & {\color[HTML]{1F1F1F} 3, 6, 10}                  \\ \hline
{\color[HTML]{1F1F1F} Gengram Ngram-sizes}         & \multicolumn{1}{c|}{{\color[HTML]{1F1F1F} \textbackslash{}}}                     & \multicolumn{2}{c|}{{\color[HTML]{1F1F1F} 1 to 6}}                                                                                 \\ \hline
{\color[HTML]{1F1F1F} Gengram Window-sizes}        & \multicolumn{1}{c|}{{\color[HTML]{1F1F1F} \textbackslash{}}}                     & \multicolumn{1}{c|}{{\color[HTML]{1F1F1F} \textbackslash{}}}                              & {\color[HTML]{1F1F1F} 21}                        \\ \hline
{\color[HTML]{1F1F1F} Gengram Conv-kernel-size}    & \multicolumn{1}{c|}{{\color[HTML]{1F1F1F} \textbackslash{}}}                     & \multicolumn{2}{c|}{{\color[HTML]{1F1F1F} 4}}                                                                                                \\ \hline
{\color[HTML]{1F1F1F} Gengram Embed-dim-per-ngram} & \multicolumn{1}{c|}{{\color[HTML]{1F1F1F} \textbackslash{}}}                     & \multicolumn{2}{c|}{{\color[HTML]{1F1F1F} 1024}}                                                                                             \\ \hline
{\color[HTML]{1F1F1F} Seq Length}                  & \multicolumn{3}{c|}{{\color[HTML]{1F1F1F} 8192 / 32768}}                                                                                                                                                                        \\ \hline
{\color[HTML]{1F1F1F} Global Batch Size}           & \multicolumn{3}{c|}{{\color[HTML]{1F1F1F} 1024}}                                                                                                                                                                                \\ \hline
{\color[HTML]{1F1F1F} Learning Rate}               & \multicolumn{3}{c|}{{\color[HTML]{1F1F1F} 1e-4 to 1e-5}}                                                                                                                                                                        \\ \hline
{\color[HTML]{1F1F1F} LR Scheduler}                & \multicolumn{3}{c|}{{\color[HTML]{1F1F1F} Cosine Decay}}                                                                                                                                                                        \\ \hline
\end{tabular}
    \end{sc}
  \end{small}
  \vskip -0.1in
\end{table}

\begin{table}[h!]
\caption{MoE-1.2B Configurations For Layer and Window}
  \label{atable:1.2b model structure}
  \centering 
  \begin{small}
    \begin{sc}
    \begin{tabular}{|l|cc|}
    \hline
    \rowcolor[HTML]{EFEFEF} 
{\color[HTML]{1F1F1F} Parameter}                   & \multicolumn{1}{c|}{\cellcolor[HTML]{EFEFEF}{\color[HTML]{1F1F1F} 1.2B-Selected-Layers}} & {\color[HTML]{1F1F1F} 1.2B-Selected-Windows}                                                                                        \\ \hline
{\color[HTML]{1F1F1F} Total Parameters}            & \multicolumn{2}{c|}{{\color[HTML]{1F1F1F} 1.2B}}                                                                                                                                                                               \\ \hline
{\color[HTML]{1F1F1F} Activated Parameters}           & \multicolumn{2}{c|}{{\color[HTML]{1F1F1F} 0.3B}}                                                                                                                                                                               \\ \hline
{\color[HTML]{1F1F1F} Hidden Size}                 & \multicolumn{2}{c|}{{\color[HTML]{1F1F1F} 1024}}                                                                                                                                                                               \\ \hline
{\color[HTML]{1F1F1F} Num Layers}                  & \multicolumn{2}{c|}{{\color[HTML]{1F1F1F} 12}}                                                                                                                                                                                 \\ \hline
{\color[HTML]{1F1F1F} Attention Heads}             & \multicolumn{2}{c|}{{\color[HTML]{1F1F1F} 16}}                                                                                                                                                                                 \\ \hline
{\color[HTML]{1F1F1F} GQA Query Groups}            & \multicolumn{2}{c|}{{\color[HTML]{1F1F1F} 8}}                                                                                                                                                                                  \\ \hline
{\color[HTML]{1F1F1F} Normalization}               & \multicolumn{2}{c|}{{\color[HTML]{1F1F1F} RMSNorm}}                                                                                                                                                                            \\ \hline
{\color[HTML]{1F1F1F} Activation}                  & \multicolumn{2}{c|}{{\color[HTML]{1F1F1F} SwiGLU}}                                                                                                                                                                             \\ \hline
{\color[HTML]{1F1F1F} Positional Emb}              & \multicolumn{2}{c|}{{\color[HTML]{1F1F1F} RoPE}}                                                                                                                                                                               \\ \hline
{\color[HTML]{1F1F1F} Num Experts}                 & \multicolumn{2}{c|}{{\color[HTML]{1F1F1F} 8}}                                                                                                                                                                                  \\ \hline
{\color[HTML]{1F1F1F} Router Top-k}                & \multicolumn{2}{c|}{{\color[HTML]{1F1F1F} 2}}                                                                                                                                                                                  \\ \hline
{\color[HTML]{1F1F1F} MoE Operators}               & \multicolumn{2}{c|}{{\color[HTML]{1F1F1F} Grouped GEMM}}                                                                                                                                                                       \\ \hline
{\color[HTML]{1F1F1F} Load Balancing}              & \multicolumn{2}{c|}{{\color[HTML]{1F1F1F} Aux Loss}}                                                                                                                                                                           \\ \hline
{\color[HTML]{1F1F1F} Gengram Layers}              & \multicolumn{1}{c|}{{\color[HTML]{1F1F1F} L $\in$ {[}1, 12{]}}}                          & {\color[HTML]{1F1F1F} 10}                                                                                                           \\ \hline
{\color[HTML]{1F1F1F} Gengram Ngram-sizes}         & \multicolumn{2}{c|}{{\color[HTML]{1F1F1F} 1 to 6}}                                                                                                                                                                   \\ \hline
{\color[HTML]{1F1F1F} Gengram Window-sizes}        & \multicolumn{1}{c|}{{\color[HTML]{1F1F1F} \textbackslash{}}}                             & {\color[HTML]{1F1F1F} \begin{tabular}[c]{@{}c@{}}S $\in$ \\ {[}6, 8,9,10,12,14,15,16,\\ 18,20,21,22,24,26,27,28,30{]}\end{tabular}} \\ \hline
{\color[HTML]{1F1F1F} Gengram Conv-kernel-size}    & \multicolumn{2}{c|}{{\color[HTML]{1F1F1F} 4}}                                                                                                                                                                                  \\ \hline
{\color[HTML]{1F1F1F} Gengram Embed-dim-per-ngram} & \multicolumn{2}{c|}{{\color[HTML]{1F1F1F} 1024}}                                                                                                                                                                               \\ \hline
{\color[HTML]{1F1F1F} Seq Length}                  & \multicolumn{2}{c|}{{\color[HTML]{1F1F1F} 8192}}                                                                                                                                                                               \\ \hline
{\color[HTML]{1F1F1F} Global Batch Size}           & \multicolumn{2}{c|}{{\color[HTML]{1F1F1F} 1024}}                                                                                                                                                                               \\ \hline
{\color[HTML]{1F1F1F} Learning Rate}               & \multicolumn{2}{c|}{{\color[HTML]{1F1F1F} 1e-4 to 1e-5}}                                                                                                                                                                       \\ \hline
{\color[HTML]{1F1F1F} LR Scheduler}                & \multicolumn{2}{c|}{{\color[HTML]{1F1F1F} Cosine Decay}}                                                                                                                                                                       \\ \hline
    \end{tabular}

    \end{sc}
  \end{small}
  \vskip -0.1in
\end{table}

\begin{table}[!h]
\caption{Load Balancing Test}
  \label{atable:10Bmodel structure}
  \centering 
  \begin{small}
  \begin{sc}
\begin{tabular}{|l|ccc|}
\hline
\rowcolor[HTML]{EFEFEF} 
{\color[HTML]{1F1F1F} Parameter}                   & \multicolumn{3}{c|}{\cellcolor[HTML]{EFEFEF}{\color[HTML]{1F1F1F} 10B MoE (Engram)}}                                \\ \hline
{\color[HTML]{1F1F1F} Total Parameters}            & \multicolumn{3}{c|}{{\color[HTML]{1F1F1F} 10B}}                                                                     \\ \hline
{\color[HTML]{1F1F1F} Activated Parameters}           & \multicolumn{1}{c|}{{\color[HTML]{1F1F1F} 0.15625B}} & \multicolumn{1}{c|}{{\color[HTML]{1F1F1F} 0.3125B}} & 0.625B \\ \hline
{\color[HTML]{1F1F1F} Hidden Size}                 & \multicolumn{3}{c|}{{\color[HTML]{1F1F1F} 1280}}                                                                    \\ \hline
{\color[HTML]{1F1F1F} Num Layers}                  & \multicolumn{3}{c|}{{\color[HTML]{1F1F1F} 12}}                                                                      \\ \hline
{\color[HTML]{1F1F1F} Attention Heads}             & \multicolumn{3}{c|}{{\color[HTML]{1F1F1F} 16}}                                                                      \\ \hline
{\color[HTML]{1F1F1F} GQA Query Groups}            & \multicolumn{3}{c|}{{\color[HTML]{1F1F1F} 8}}                                                                       \\ \hline
{\color[HTML]{1F1F1F} Normalization}               & \multicolumn{3}{c|}{{\color[HTML]{1F1F1F} RMSNorm}}                                                                 \\ \hline
{\color[HTML]{1F1F1F} Activation}                  & \multicolumn{3}{c|}{{\color[HTML]{1F1F1F} SwiGLU}}                                                                  \\ \hline
{\color[HTML]{1F1F1F} Positional Emb}              & \multicolumn{3}{c|}{{\color[HTML]{1F1F1F} RoPE}}                                                                    \\ \hline
{\color[HTML]{1F1F1F} Num Experts}                 & \multicolumn{1}{c|}{{\color[HTML]{1F1F1F} 128}}      & \multicolumn{1}{c|}{{\color[HTML]{1F1F1F} 64}}      & 32     \\ \hline
{\color[HTML]{1F1F1F} Router Top-k}                & \multicolumn{3}{c|}{{\color[HTML]{1F1F1F} 2}}                                                                       \\ \hline
{\color[HTML]{1F1F1F} Load Balancing}              & \multicolumn{3}{c|}{{\color[HTML]{1F1F1F} Aux Loss}}                                                                \\ \hline
{\color[HTML]{1F1F1F} Gengram Layers}              & \multicolumn{3}{c|}{{\color[HTML]{1F1F1F} \{3, 6, 10\}}}                                                            \\ \hline
{\color[HTML]{1F1F1F} Gengram Ngram-sizes}         & \multicolumn{3}{c|}{{\color[HTML]{1F1F1F} 1 to 6}}                                                                \\ \hline
{\color[HTML]{1F1F1F} Gengram Window-sizes}        & \multicolumn{3}{c|}{{\color[HTML]{1F1F1F} /}}                                                                       \\ \hline
{\color[HTML]{1F1F1F} Gengram Conv-kernel-size}    & \multicolumn{3}{c|}{{\color[HTML]{1F1F1F} 4}}                                                                       \\ \hline
{\color[HTML]{1F1F1F} Gengram Embed-dim-per-ngram} & \multicolumn{3}{c|}{{\color[HTML]{1F1F1F} 1024}}                                                                    \\ \hline
{\color[HTML]{1F1F1F} Seq Length}                  & \multicolumn{3}{c|}{{\color[HTML]{1F1F1F} 8192}}                                                                    \\ \hline
{\color[HTML]{1F1F1F} Global Batch Size}           & \multicolumn{3}{c|}{{\color[HTML]{1F1F1F} 1024}}                                                                    \\ \hline
{\color[HTML]{1F1F1F} Learning Rate}               & \multicolumn{3}{c|}{{\color[HTML]{1F1F1F} 1e-4 to 1e-5}}                                                            \\ \hline
{\color[HTML]{1F1F1F} LR Scheduler}                & \multicolumn{3}{c|}{{\color[HTML]{1F1F1F} Cosine Decay}}                                                            \\ \hline
\end{tabular}
    \end{sc}
  \end{small}
  \vskip -0.1in
\end{table}


\FloatBarrier

\section{Biological interpretability of the hash table}
We further probe the interpretability of the 3-gram hash table learned by Gengram. Because DNA is double-stranded, signals tied to structure/accessibility are expected to be approximately invariant under reverse complementation (RC). Meanwhile, if the table behaves like a motif dictionary, codon representations would tend to collapse by amino-acid semantics (i.e., synonymous codons cluster together). We therefore examine the hash table from two complementary angles: RC symmetry and synonymous-codon clustering consistency, to determine whether it functions as a reusable motif memory rather than a static motif dictionary.

For each layer, we enumerate all 64 codons and construct the reverse complement $rc(c)$ for each codon $c$. We compute the cosine distance $d(c, rc(c))$ in representation space. Using a cluster map obtained by clustering codon embeddings, we categorize each RC pair as:

\textbf{within:} $c$ and $rc(c)$ fall into the same cluster, so $d(c,rc(c))$ contributes to the within distribution;

\textbf{between:} otherwise, it contributes to the between distribution.

We compare within vs. between using a Mann–Whitney U test (as implemented in your script and visualized by the violin plots). If RC equivalence is learned, RC pairs should be more “cluster-aligned,” i.e., within distances become significantly smaller than between.

\begin{figure*}[ht]
  \vskip 0.2in
  \begin{center}
    \centerline{\includegraphics[width=0.8\textwidth]{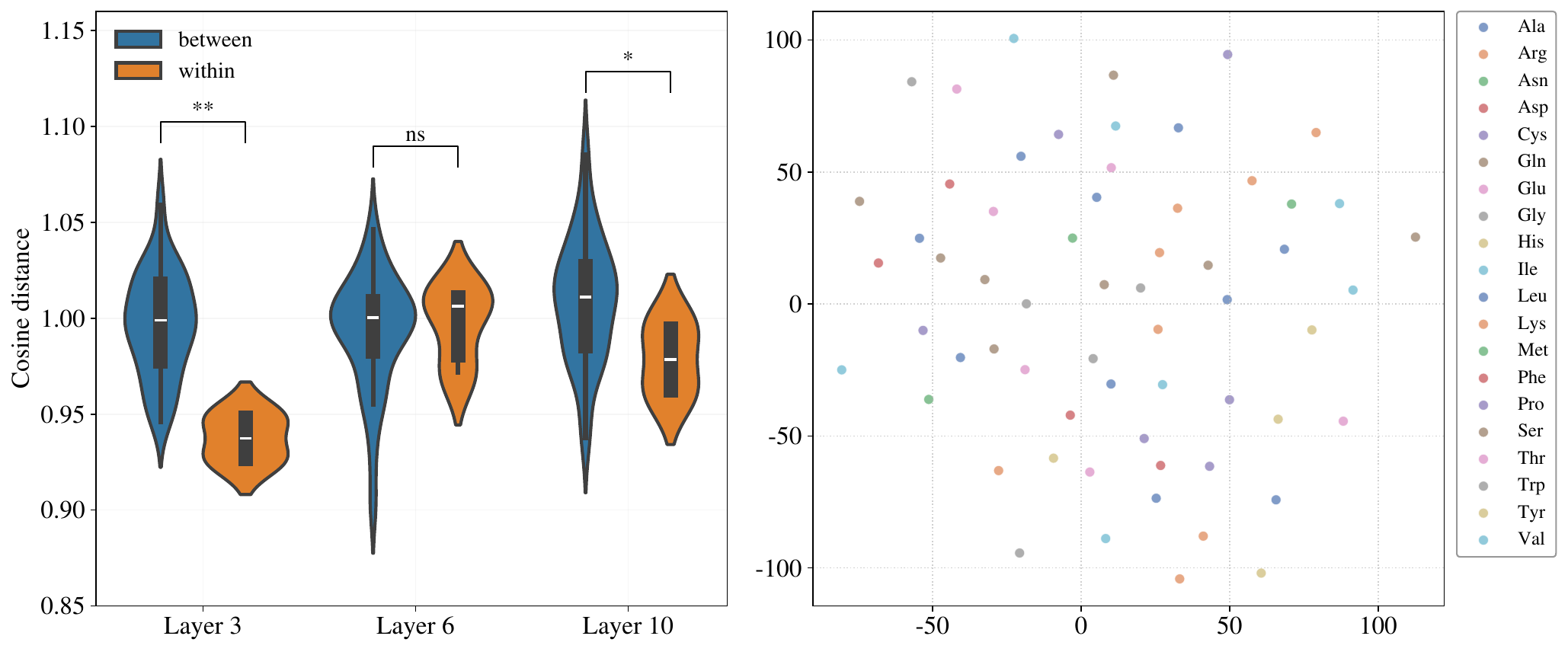}}
    \caption{
      \textbf{RC violin (within vs between) at left:} demonstrates whether RC pairs are aligned in the learned memory space. The key takeaway is: smaller within $\Rightarrow$ RC pairs tend to co-cluster and be more similar $\Rightarrow$ memory captures strand/double-helix equivalence, with a meaningful layer-wise specialization (significant at L3/L10 but not at L6).
        \textbf{t-SNE by AA at right:} serves as a counter-evidence showing the representations do not trivially cluster by amino-acid identity, supporting “memory rather than dictionary.”
   }
    \label{fig:violin}
  \end{center}
\end{figure*}

The RC\-within vs RC\-between comparison reveals a clear layer-wise pattern. In Layer\-3 and Layer\-10, within distances are generally smaller than between distances, with statistically significant separation (marked as ** and * in the Figure \ref{fig:violin}) left, while in Layer\-6 the difference is not significant (ns). This suggests that the 3-gram memory does not enforce RC equivalence uniformly across layers; instead, it exhibits a division of labor where early or deep layers are more RC-aligned, whereas the middle layer is more task-sensitive. This is biologically plausible: intermediate representations may emphasize strand-aware cues such as transcription directionality and splice donor/acceptor patterns (which are inherently not RC-invariant), while deeper layers integrate broader context and favor features tied to double-stranded structure/accessibility, leading to stronger RC pairing alignment (more “same-cluster” RC pairs and smaller within distances).

We also find that synonymous codons do not systematically collapse into the same cluster. For many amino acids, their synonymous codons are distributed across multiple clusters; quantitatively, the agreement between cluster labels and amino-acid labels is weak (e.g., near-zero ARI and only moderate NMI). This behavior argues against the hash table being a static motif dictionary that compresses codons purely by amino-acid semantics. Instead, it is more consistent with a motif memory: representations jointly reflect codon usage bias, local sequence context, coupling with regulatory constraints (e.g., splice-related short motifs), and mixed coding/non-coding pressures, hence they are not fully governed by the “synonymous = equivalent” rule (Figure \ref{fig:violin} right).

\end{document}